\documentclass{revtex4}

\newcommand{\beq}{\begin{equation}}  
\newcommand{\eeq}{\end{equation}}
\newcommand{\bea}{\begin{eqnarray}}
\newcommand{\eea}{\end{eqnarray}}

\usepackage{graphicx}

\begin{document}

\title{The minimal supersymmetric grand unified theory: \\
I. symmetry breaking and the particle spectrum}

\author{
Borut Bajc$^{(1)}$,  
Alejandra Melfo$^{(2)}$, Goran Senjanovi\'c$^{(3)}$ 
and Francesco Vissani$^{(4)}$}
\affiliation{$^{(1)}$ {\it J.\ Stefan Institute, 1001 Ljubljana, Slovenia}}
\affiliation{$^{(2)}$ {\it Centro de F\'{\i}sica Fundamental, 
Universidad de Los Andes, M\'erida, Venezuela}}
\affiliation{$^{(3)}${\it International Center for Theoretical Physics, 
Trieste, Italy}}
\affiliation{$^{(4)}${\it INFN, Laboratori Nazionali 
del Gran Sasso, Theory Group, Italy}}

\begin{abstract}

 We discuss in detail the symmetry breaking and related issues in the
minimal renormalizable supersymmetric grand unified theory. We
compute the particle spectrum and study its impact on the physical
scales of the theory. This provides a framework for the analysis of
phenomenological implications of the theory, to be carried out in 
part II. 

\end{abstract}
\maketitle

\section{Introduction}\hspace{0.5cm} 

It has been argued recently \cite{Aulakh:2003kg} 
that the minimal supersymmetric
renormalizable grand unified theory is based on the $SO(10)$ gauge
symmetry with the following minimal set of states \cite{mc,Clark:ai}
\begin{itemize}
\item three generations of 16-dimensional matter supermultiplets
$16_F$

\item $ 210_H$, $126_H$, $\bar{126_H}$ and $10_H$ Higgs supermultiplets.

\end{itemize}

The theory is minimal in the sense of having a minimal set of
parameters and most predictability. Its main features are:

\begin{enumerate}
\item exact R-parity conservation at all energies \cite{Aulakh:1997ba,
Aulakh:1998nn,Aulakh:1999cd}
and a stable LSP

\item natural smallness of neutrino masses through the see-saw
mechanism \cite{yana,glashow,gmrs,Mohapatra:1979ia}

\item completely realistic fermion spectrum 
\cite{Babu:1992ia,Oda:1998na,Brahmachari:1997cq}

\item in the case of type II see-saw, it offers a natural con connection
between $b-\tau$ unification and a large atmospheric neutrino mixing 
angle \cite{Bajc:2002iw}

\item both in the type I and type II see-saw cases, the 1-3 leptonic
mixing angle turns out to be large, close to the upper experimental
limit \cite{Matsuda:2001bg,Goh:2003sy,Goh:2003hf} 

\item the loss of asymptotic freedom above $M_{GUT}$ and the existence
of a new fundamental scale $M_F \simeq 10 M_{GUT}$ where couplings
become strong \cite{c}

\end{enumerate}

This theory thus should be confronted with experiment, which requires
the detailed computation of the symmetry breaking. This is the aim of
this paper. In the follow-up, we will discuss at length the
phenomenological implications: proton decay, neutrino masses and
mixings, R-parity, leptogenesis and related issues.

Some initial attempts in this direction have already been made 
\cite{Lee:1993rr,charanaarti}, but the complete analysis requires a more 
precise information about the particle states.

The paper is organized as follows. In the next section, we describe
the construction of the theory and argue in favor of its
minimality. In section III, we study the patterns of symmetry breaking
allowed by the most general renormalizable superpotential. The spectrum
is accordingly computed in section IV, and in section V we use it to
determine the physical scales. Summary and outlook is left for the 
last section. 
Many technical details and useful tables are left for the Appendices.

\section{The theory: fields and interactions}

As in any $SO(10)$ theory, the matter superfields are
16-dimensional spinorial representations. The Higgs sector
\cite{mc,Clark:ai,Lee:1993rr} contains as mentioned
\beq
\Phi (210) ; \quad \Sigma (126) ; 
\quad \overline{\Sigma} (\overline{126}) ; \quad H (10)
\eeq

$\bar{\Sigma}$ is needed in order to give a large mass $M_R$ to $\nu^c$, 
$\Sigma$ in order to preserve supersymmetry at $M_R$ and 
$\Phi (210)$ in order to complete the symmetry breaking down to 
MSSM. $\Phi$ is the minimal choice that does the job and it also plays 
an important role in generating the correct fermion mass matrices 
(Section V).  Having  realistic fermionic spectrum necessitates also 
$H$ on top of $\bar \Sigma$.

The most general renormalizable superpotential of the above fields is 

\begin{eqnarray}
W_H &=& \frac{m_\Phi}{4!}\Phi^2
+\frac{m_\Sigma}{5!}\Sigma\overline\Sigma + 
\frac{\lambda}{4!}\Phi^3 +
\frac{\eta}{4!}\Phi \Sigma \overline\Sigma \nonumber \\
&+& m_H H^2 + \frac{1}{4!} \Phi H (\alpha
\Sigma + \bar\alpha \overline{\Sigma})
\label{superpot}
\end{eqnarray}

The simplicity of $W_H$ is worth commenting on. It has only four
different couplings  and three mass terms, which facilitates the study
of symmetry breaking. This is the advantage of large representations,
which may have other defects. 

The matter superfields are $\Psi_a (16)$, $a=1,2,3$. The most general
Yukawa superpotential is given by

\beq
W_Y = \Psi (Y_{H} H + Y_{\Sigma} \overline\Sigma) \Psi\;.
\label{yukawa}
\eeq
with generation indices suppressed, $Y_H$ and $Y_\Sigma$  are
symmetric matrices. This implies 15 real  couplings in total. Namely, $Y_H$
for example can be diagonalized and made real, which means 3 real
couplings and $Y_\Sigma$ has 6 complex or 12 real couplings.

The small number of couplings should be considered the main virtue of
the theory, for it implies a large amount of predictivity. After
rotating away the phases of Higgs superfields, the 7 (4 couplings and
3 masses) complex parameters of $W_H$ become 10 real ones. Together
with the single gauge coupling, we have thus 26 real parameters in
total.  Of course, we are not counting the supersymmetry breaking
terms. This can be compared with the MSSM, which has the same number
of couplings but describes far less phenomena.
Similarly, one can show that the $SU(5)$ supersymmetric theory has
many more couplings \cite{Aulakh:2003kg}. As we have argued at 
length in \cite{Aulakh:2003kg},
this theory should be considered the minimal supersymmetric GUT.

The study of symmetry breaking and fermion masses favor the Pati-Salam
$G_{422}=  SU(4)_C\times SU(2)_L\times SU(2)_R$ language \cite{charanaarti}. The decomposition of
the above fields under $G_{442}$ is given by

\begin{eqnarray}
H \equiv {\bf 10} & = & (6,1,1) + (1,2,2) \nonumber \\
\Psi \equiv {\bf 16} &=& (4,2,1) +(\bar{4},1,2) \nonumber \\
\Phi \equiv {\bf 210} & = & (15,1,1) + (1,1,1) + (15,1,3) \nonumber \\
& + &(15,3,1) + (6,2,2) + (10,2,2) + (\overline{10},2,2) \nonumber \\
\Sigma \equiv  {\bf 126} & = & (\overline{10},1,3) + 
(10,3,1) + (6,1,1) + (15,2,2) \nonumber \\
\overline\Sigma \equiv \overline{{\bf 126}} & = & (10,1,3) + 
(\overline{10},3,1) + (6,1,1) + (15,2,2) \nonumber 
\end{eqnarray} 

Of course, in order to study in detail 
the symmetry breaking  and the particle spectrum, a complete
decomposition under the Standard Model group of the above fields is
required. We refer the reader to Appendix \ref{one}
for these details.

\section{Patterns of symmetry breaking}

The first step is the breaking of $SO(10)$ down to the MSSM; and here
$H$ can be ignored. Only the MSSM singlets are allowed to take a
vacuum expectation value (VEV). We shall call their VEVs
\begin{eqnarray}
p = \langle\Phi(1,1,1)\rangle ; \quad a = \langle  
\Phi(1,1,15)\rangle ; \quad
\omega = \langle \Phi(1,3,15)\rangle \nonumber \\
\sigma = \langle \Sigma(1,3,\overline{10})\rangle ; \quad 
\bar\sigma = \langle \bar{\Sigma}(1,3,10)\rangle
\end{eqnarray}

The superpotential as a 
function of these VEVs is calculated to be

\begin{eqnarray}
W_H&=&m_\Phi\left(p^2+3a^2+6\omega^2\right)
+2 \lambda\left(a^3+3p\omega^2+6a\omega^2\right)\nonumber\\
&+&m_\Sigma \sigma\bar\sigma+\eta\sigma\bar\sigma\left(
p+3a - 6 \omega\right)\;.
\end{eqnarray}

Vanishing of the D-terms implies $|\sigma|=|\bar\sigma|$, while from the
F-terms we get

\begin{eqnarray}
2\,m_\Phi p+6\,\lambda \omega^2+\eta\sigma\bar\sigma &=& 0 \\
2\,m_\Phi a+2\,\lambda (a^2+2\omega^2)+\eta\sigma\bar\sigma &=& 0 \\
2\,m_\Phi\omega +2\,\lambda (p+2a)\omega +\eta\sigma\bar\sigma &=&0 \\
\sigma\left[ m_\Sigma + \eta ( p + 3 a - 6 \omega )\right] &=& 0 
\end{eqnarray}

{}From these equations, we get
 the following set of degenerate SUSY-preserving vacua
\begin{enumerate}
\item $ p = a = \omega =\sigma =0$ --the $SO(10)$-preserving minimum.

\item $p = a = -\omega = -m_\Phi/3\lambda$ ; $\sigma = 0$. As can be
confirmed by calculating explicitly the gauge boson masses, this
minimum has $SU(5)\times U(1)$ symmetry.

\item $p = a = -\omega = -m_\Sigma/10 \eta $ ; $\sigma\bar\sigma  = 
m_\Sigma (10\eta m_\Phi - 3
\lambda m_\Sigma)/(50 \eta^3)$. This is the $SU(5)$ minimum, and 
includes the previous one for   $\lambda m_\Sigma/\eta m_\Phi = 10/3$

\item $ p = \omega = \sigma=0$ ; $ a = -m_\Phi/\lambda$. This is obviously
the left-right symmetric $SU(3)_C\times SU(2)_L\times SU(2)_R 
\times U(1)_{B-L}$ minimum. 

\item $p = a = \omega = -m_\Phi/3\lambda$ ; $\sigma = 0$. This is again
$SU(5)\times U(1)$ symmetric, but    with the flipped $SU(5)$
assignments for the particle states. 

\item $p = 3 m_\Phi/\lambda$ ; $a = -2 m_\Phi/\lambda$, $\omega = \pm i
m_\Phi/\lambda$ ; $\sigma =0$. This minimum has symmetry
 $SU(3)_C\times SU(2)_L\times U(1)_R \times U(1)_{B-L}$. 

\item
\beq
\label{7vevs}
p =-\frac{m_\Phi}{\lambda} \frac{x (1-5x^2)}{(1 - x)^2}\;;\;
a=-\frac{m_\Phi}{\lambda} \frac{(1 - 2x - x^2)}{(1-x)}\;;\;
\omega = - \frac{m_\Phi}{\lambda}x \;;\; 
\sigma\bar\sigma =\frac{2m_\Phi^2}{\eta\lambda} 
\frac{x (1 -3 x)(1 + x^2)}{(1 - x)^2} 
\eeq

\beq 
\label{7g}
-8 x^3 + 15 x^2 - 14 x + 3 = (x-1)^2 \frac{\lambda m_\Sigma}{\eta
m_\Phi}
\eeq
\end{enumerate}

For generic $x$, this is the Standard Model minimum.
Of particular interest are the cases $x\sim 0$ and $x\sim 1$, which
provide the chains with intermediate scales. The former case
corresponds to the left-right symmetry, while the latter gives an
intermediate $G_{422}$ (Pati-Salam) scale. 
  
This  solution includes:
\begin{itemize}
\item the 3rd (for $x = 1/2$), if $  \lambda m_\Sigma/\eta m_\Phi = -5$

\item the 4th (for $x=0$), if $ \lambda m_\Sigma/\eta m_\Phi = 3$

\item the 5th  (for $x = 1/3$), if $\lambda m_\Sigma/\eta m_\Phi = -2/3$

\item the 6th   (for $x=\pm i$), if 
$\lambda m_\Sigma/\eta m_\Phi = -3(1\pm 2i)$
\end{itemize}

All this will become more transparent when we calculate the particle
spectrum in the following section.

\section{Particle spectrum}

We refer the reader to Appendix \ref{one} for the notation and 
decomposition in SM language, and to Appendix \ref{two} for 
details of the calculation of the spectrum. The computations are 
performed using a nice method developed by He and Meljanac \cite{He:jw}. 
We are interested in the physically realistic case 7, where the 
resulting symmetry is the MSSM. The unmixed states are given in Table 
\ref{table1}.

\begin{table}
$$
\begin{array}{|c|l|c|}
\hline
{\rm Field} & SU(3)_C\times SU(2)_L\times U(1)_Y & {\rm Mass}/m_\Phi \\
\hline
\Phi &(3,1,+ 5/3),(\bar 3,1,- 5/3)  & 
8\,x \left( 2\,x-1 \right)  /  \left( x-1 \right) ^{2} \\
&(8,1,\pm 1) & 
4\,({x}^{2}-3\,x+1+3\,{x}^{3}) /  \left( x-1 \right) ^{2} \\
&(1,3,0) &
-2\,({x}^{2}-5\,x+1+7\,{x}^{3}) /  \left( x-1 \right) ^{2} \\
&(3,3,- 2/3),(\bar 3,3,+ 2/3)&
-4\,x \left( -1+3\,{x}^{2} \right)  /  \left( x-1 \right) ^{2}
 \\
&(8,3,0) & 
-4\,(-{x}^{2}+2\,x-1+2\,{x}^{3}) /  \left( x-1 \right) ^{2} \\
&(1,2,\pm 3/2) & 
-4\,(-1+x+3\,{x}^{2}) / \left( x-1 \right)  \\
&(6,2,- 1/6),(\bar 6,2,+ 1/6)&
4\,(-1+x+{x}^{2}) / \left( x-1 \right)  \\
&(6,2,+ 5/6), (\bar 6,2, - 5/6)&
4\,(2\,x-1) / \left( x-1 \right)  \\
\hline
\Sigma,\bar{\Sigma} 
&(1,3,-1),(1,3, +1)&
 - 4\,\left(\eta/\lambda\right)\, x \left( 4\,{x}^{2}-3\,x+1 \right) 
  /  \left( x-1 \right) ^{2} \\
& (3,3, - 1/3),(\bar 3,3,+ 1/3)  &
- 2\,\left(\eta/\lambda\right)\, \left( 7\,{x}^{3}- 7\,{x}^{2}+ 5\,x- 1
 \right)  /  \left( x-1 \right) ^{2}\\
& (6,3,+ 1/3),(\bar 6,3,- 1/3)  & 
- 4\, \left(\eta/\lambda\right)\,  \left( 3\,x -1 \right)  
\left( {x}^{2}-x+1 \right) 
  /  \left( x-1 \right) ^{2} \\
& (1,1, +2),(1,1, -2) &
-12\,x\left(\eta/\lambda\right) \\
& (\bar 3,1,+4/3),(3,1, -4/3)  &
-2 \, \left(\eta/\lambda\right)\, \left( 3\,{x}^{2}- 6 \,x+ 1 \right)  / 
\left( x-1 \right) 
\\
& (\bar 6,1,+ 2/3), (6,1,- 2/3)  &
-4 \left(\eta/\lambda\right)\, \left( 1-3\,x \right)  / 
\left( x-1 \right) \\
& (\bar 6,1,- 1/3), (6,1,+ 1/3)  &
-2\left(\eta/\lambda\right)\, \left( {x}^{2}- 7\,x+ 2 \right)  
/ \left( x-1 \right)
\\
& (\bar 6,1,- 4/3), (6,1, + 4/3)& 
-4 \left(\eta/\lambda\right)\, \left( {x}^{2}- 4\,x+ 1 \right)  
/ \left( x-1 \right)
\\
& (3,2, + 7/6),(\bar 3,2,- 7/6) & 
- 2\,\left(\eta/\lambda\right)\, \left( 6\,{x}^{3}-10\,{x}^{2}+7\,x-1
 \right)  /  \left( x-1 \right) ^{2} \\
& (\bar 3,2, -1/6),(3,2, +1/6)  &
- 2\,  \left(\eta/\lambda\right)\, \left( 4\,{x}^{3}- 6\,{x}^{2}+ 5\,x- 1
 \right)  /  \left( x-1 \right) ^{2}\\
& (\bar 3,2,- 7/6), ( 3,2, + 7/6)  &
 - 2\,\left(\eta/\lambda\right)\, \left( 5\,{x}^{3}- 8\,{x}^{2}+ 6\,x-1
 \right)  /  \left( x-1 \right) ^{2}\\
& (8,2,+ 1/2),(8,2,- 1/2) & 
- 2\,\left(\eta/\lambda\right)\, \left( 3\,{x}^{3}- 7\,{x}^{2}+ 8\,x- 2
 \right)  /  \left(x -1 \right) ^{2} \\
& (8,2, - 1/2),(8,2, + 1/2)&
- 2\, \left(\eta/\lambda\right)\, \left( 4\,{x}^{3}- 9\,{x}^{2}+ 9\,x- 2
 \right)  /  \left( x-1 \right) ^{2} \\
\hline
\end{array}
$$
\caption{ Masses of the unmixed states as functions of $x$ for the 7th
symmetry breaking pattern. \label{table1}}
\end{table}

We  give the mixing matrices for the rest of the states in 
Appendix \ref{three}. 
In order to find the eigenvalues of a general matrix $M$ one needs to 
diagonalize the matrix $M^\dagger M$. Since our aim here is not the 
complete numerical and phenomenological analysis (left for the part II), 
we give for illustration the eigenvalues  in 
the Hermitian case (Table \ref{table2}). This case by definition means 
the CP conserving situation at the high scale of real couplings and 
real vevs and $\bar\sigma=\sigma$ and $\bar\alpha=\alpha$. 

\begin{table}
$$
\begin{array}{|c|l|c|}
\hline
{\rm Fields}  & SU(3)_C\times SU(2)_L\times U(1)_Y & {\rm Mass}/m_\Phi \\
\hline
 \Phi,\Sigma\bar\Sigma & (3,1,+2/3),(\bar 3,1,-2/3) & 
\begin{array}{c}
0 \\
x(3 x^2 -1)/(x-1)^2 -(\eta/\lambda)  ({x}^{2}-4\,x+1)/(x-1) 
  \pm  \sqrt{A} \, ; \\
A= \left[ x(3 x^2 -1)/(x-1)^2 +(\eta/\lambda)
({x}^{2}-4\,x+1)/(x-1)\right]^2 \\
-4(3\,x -1 ) ( 1+{x}^{2} )  ( 
-2\,\lambda\,x+3\,\eta\,x-3\,\eta ) x /[ \lambda\, (x -1)^{3}]
\end{array} \\ \hline
 \Phi,\Sigma\bar\Sigma & (1,1,\pm 1) & 
\begin{array}{c}
0 \\
  2  (3\,x -1 )(1 + x^2) /
 ( x -1 )^2  \,- \, 6\, x \,  \eta /\lambda     
\end{array} \\ \hline
\Phi & (8,1,0) & 
\begin{array}{c}
\left[ x\,(7 x + 11)\,(x-1)\,+\,4 \,\pm  \, x \sqrt{B}\, \right ] 
/ \left(x -1 \right) ^{2};
 \\
B= 32\, x\, (x^2+1)\, (x-1) \,+\, (x^2 +3 )^2
\end{array} \\ \hline
\Phi,\Sigma\bar\Sigma & (3,2,+1/6),(\bar 3,2,-1/6)  & 
\begin{array}{c}
0 \\
(x^2-2 x -1)/(x-1) +(\eta/\lambda) (3 x -1)(x^2 -x +1)/(x-1)^2  
\pm \sqrt{C}\,  ;\\
C= [(x^2-2 x -1)/(x-1) - (\eta/\lambda)(3 x -1)(x^2 -x +1)/(x-1)^2 ]^2
\\
+ 4 x (x^2 +1)\left[ 2 - (\eta/\lambda) 3(3 x -1)/(x-1)  \right]/(x-1)
\end{array} \\ \hline 
\Phi,\Sigma\bar\Sigma & (3,2,-5/6),(3,2,+5/6)  & 
\begin{array}{c}
0 \\
-2 (1-2\,x+3\,{x}^{2})/(x-1) 
\end{array} \\ \hline
\Phi,\Sigma\bar\Sigma & (1,1,1) & 
\begin{array}{c}
0 \\
\{Z_1,Z_2,Z_3,Z_4 \}/(\lambda(x-1)) \,, \;{\rm with}\,  Z_i \, {\rm
roots\,  of } \\
 128\,x\eta\, \left(3\,x -1 \right)  \left(2\,x -1 \right)  \left( 1+
x \right)  \left( 1+{x}^{2} \right)  \left( {x}^{3}-3\,{x}^{2}+2\,x-1
 \right) \\
+ 8\left[  \lambda (x -1 )( 2\,x -1)( 1+x)
(6\,{x}^{4}-13\,{x}^{3}+3\,{x}^{2}-5\,x+1) \right. \\+\left. 
 2 \eta x ( 3\,x -1 ) ( 1+{x}^{2} )  ( 19\,{x}^
{3}-35\,{x}^{2}+7\,x+1 )  \right]Z
 \\
+ 4 \left[ (-20 x^5 + 36 x^4 -45 x^3 + 15 x^2 -3 x
+1)\lambda\right. \\ \left.  +
10 \eta x (x-1)(3x -1) (x^2 + 1) \right] Z^2 
-2\,\lambda\, \left( 5\,x+{x}^{3}-1-{x}^{2} \right) Z^3 \\
+ \lambda\, \left( x-1 \right) Z^4 =0 
\end{array} \\
\hline
\end{array}
$$
\caption{Masses of the states that mix as functions of $x$ for the
7th. symmetry breaking pattern, except those of states that mix with 
$H(10)$. This is valid only for the case of Hermitian mass matrices.\label{table2}}
\end{table}

The mass matrices for  the color triplets and the SM doublets that mix with 
$H (10)$ were first calculated in \cite{charanaarti} using a different
method. We have
confirmed their results, and the complete set of matrices is given in 
 Appendix \ref{three},
 where we also discuss
their eigenvalues.  
Let us briefly discuss the physical aspects of these systems. First the
$SU(2)$ doublets. Without
fine-tuning all of them have large masses, on the order of the GUT scale.
With the minimal fine-tuning one ends up with just one light pair, as
is illustrated in Appendix \ref{three}.
What is crucial though is that
they all mix, and thus the light doublets are a mixture of the original
ones. In other words, they all have nonvanishing VEVs. In particular,
in the Pati-Salam language, both $(2,2,1)$ and $(2,2,15)$ fields will
contribute to fermion masses as 
in the Georgi-Jarlskog scenario \cite{Georgi:1979df,Aulakh:2003kg}. This
is why one ends up with a realistic matter spectrum in this theory. 
Notice that the $\Phi$  plays a central role in this through the $\alpha$
and $\bar\alpha$ couplings in (\ref{superpot}): without it one
would have only $H$ or $\bar\Sigma$ give masses to fermions which 
cannot be realistic for all three generations. 

Now, the color triplet superfields. They are responsible, either 
directly or indirectly for d=5 proton decay 
\cite{Sakai:1981pk,Weinberg:1981wj}, a main threat to any 
supersymmetric grand unified theory. It is well known that for 
generic values of their masses of the order of the GUT scale, 
proton decay becomes typically too fast 
\cite{Hisano:1992jj,Lucas:1996bc,Goto:1998qg}. 
There is some uncertainty in this due to unknown fermion and 
sfermion mixing angles 
\cite{Bajc:2002bv,Bajc:2002pg,Emmanuel-Costa:2003pu} 
(see however \cite{Murayama:2001ur}), but to be on the safe 
side it would be desirable to have these states weigh more 
\cite{Goh:2003nv}.

A useful test of symmetry breaking is the identification 
of the would-have-been Goldstone bosons. When $SO(10)$ 
is broken down to $G_{321}$, 33 gauge bosons become
massive (here and in what follows, we give values of Y/2
when specifying the SM group $G_{321}$ quantum numbers):

\begin{description} 
\item [{\it i)}]( $X$, $Y$),  mediators of proton decay,
with $G_{321}=(3,2,-5/6)$ and $(\bar 3,2, 5/6)$
\item [{\it ii)}] ($X'$, $Y'$), also mediators of proton decay,
with $G_{321}=(3,2,1/6)$ and $(\bar 3,2, -1/6)$
\item [{\it iii)}] ($X_{PS}$) the $G_{422}$ leptoquarks,
 responsible for rare decays,
with $G_{321}=(3,1,2/3)$ and $(\bar 3,1,-2/3)$
\item [{\it iv)}] ($\vec W_R$), the mirror image of the 
$\vec W$ bosons, with $G_{321}=(1,1,\pm 1)$ and $(1,1,0)$
\end{description}

A quick glance at Table \ref{table2} shows that the massless states have
precisely the above quantum numbers. 

\section{Intermediate scales?}

An important issue in supersymmetric unification is the possible 
existence of intermediate scales. Of course, the success of the 
MSSM couplings unification 
\cite{Dimopoulos:1981yj,Ibanez:yh,Einhorn:1981sx,Marciano:1981un} 
favors a single step breaking, and the intermediate scales cannot 
be too far from the GUT scale.

In the case of a single intermediate scale, the  small uncertainty in 
the values of the couplings at $M_Z$ tells us that $M_I$ can be at
most an order or two of magnitude away from $M_X$. It is useful to 
know, though, whether the GUT scale gets raised or lowered in this 
instance, and we will address the issue. A more interesting situation 
emerges with two intermediate scales, since this gives us more 
freedom. Namely, after eliminating the unification coupling at the 
GUT scale, one has only two equations  for three unknowns, one may 
in principle end up with interesting new physical scales.
We first discuss the simpler case of single intermediate scale.

\subsection{One intermediate scale}

The seventh (general) pattern of symmetry breaking allows a 
discussion of intermediate scales as a particular choice of 
$x$.  For example, $x \sim 0$ means only $a$ nonvanishing, or 
a L-R symmetric step, while $x \sim 1 $ implies a Pati-Salam 
intermediate scale. Let us discuss these two cases separately.

\begin{description}

\item[\em{ i)}] $x \sim 0  $. In this limit, 
$  ( p , \omega,\sigma) << a   $ leaves only 
$G_{3221} \equiv SU(3)_C\times SU(2)_L\times SU(2)_R\times 
U(1)_{B-L}$,  the well known parity conserving extension of the 
SM, well fit for the understanding of neutrino masses. Most of the 
states get a mass $M_X = a$, except for those that get and 
intermediate scale mass $M_I \sim  (p, \omega, \sigma ) $. 
The measure of the intermediate scale is precisely 
$x \propto M_I/M_X$. Because of left-right symmetry, the 
following $G_{3221}$ states in $\Sigma$ must  have mass
$M_I$:
$$\Sigma : (1,3,1;\pm 1)  , (1,1,3;\pm 1).$$
 There are however  other
states, belonging to the  $\Phi$ multiplet, that have an intermediate
scale mass:
 $$\Phi: (3,3,1;-2/3),(\bar 3,3,1;2/3), (3,1,3;-2/3) , (\bar
3,1,3,2/3).$$
It is straightforward in this case to show that the
lower the intermediate scale is, the lower the GUT scale becomes. This
is very bad for the already existing problem of $D=5$ proton decay,
and should be discarded.

\item[\em{ i)}] $x \sim 1 $. In this case, $p = M_X$ is the larger
scale,
 while $a \sim \sigma \sim M_I$, and $\omega$ is of order
$M_I^2/M_X$. This is the case of the $G_{422} = SU(4)_C\times
SU(2)_L\times SU(2)_R$ Pati-Salam intermediate scale, of great
phenomenological interest. States of $G_{422}$
 with an intermediate scale mass $M_I$ are
$$\Phi : (15, 1, 1) , (10, 2, 2) , (\bar{10} ,2 ,2 ) $$
$$ \Sigma,\bar\Sigma : (\bar 10, 1, 3) , (10, 1, 3)  $$
However, some
states in $\Sigma(\bar 10, 1, 3) $ $\bar \Sigma(\bar 10, 1, 3) $
are lighter, with a mass $\sim M_I^2/M_X$. These are  doubly charged color
singlets $$\delta^{++} , \bar\delta^{--},$$
often appearing in supersymmetric left-right unification 
\cite{Aulakh:1998nn,Chacko:1998jz}.

As in the left-right case it can
be easily shown that the GUT scale gets lowered and we discard this
case too.
\end{description}  

\subsection{Two intermediate scales}

We have seen that $x\sim 1$ allows for the $G_{422}$ intermediate
scale. From eq. \ref{7vevs}, it follows that $\sigma$ can be made
arbitrarily small by taking $\eta \gg \lambda$. In such case, one ends
up with the hierarchy $p\gg a \gg\sigma,\omega$. This means a three-step
breaking
\beq
SO(10) \to  
SU(4)_C\times SU(2)_L\times SU(2)_R \to 
SU(3)_C\times SU(2)_L\times SU(2)_R \times U(1)_{B-L} \to 
 MSSM
\eeq

Two possibilities arise:
\begin{description}
\item[{\bf 1}] $\sigma > \omega$, with one-step  breaking of 
$SU(2)_R\times U(1)_{B-L}$.

\item[{\bf 2}] $\omega > \sigma$, the case with an intermediate 
(the third) $B-L$ symmetry.  $SU(2)_R \times U(1)_{B-L}$ gets first 
broken down to $U(1)_R\times U(1)_{B-L}$, and $\sigma$ later 
completes the breaking. 
\end{description}

Let us examine the first case. 
Calling $p=M_X, a=M_{PS},
\sigma=M_R$, the particles that run and their masses are given in
Table \ref{run}.

\begin{table}
\begin{center}
\begin{tabular}{|l|l|c|c|c|}
\hline
$G_{422}$ State &  Mass &$ b_1$ & $b_2$ & $b_3$ \\
\hline
$\begin{array}{l}
H(6,1,1)\\
\Phi(15,1,1),{\rm except}\\
\;\;\;\;\;{\rm  \, for \, color \, octets} \\
\Sigma(\bar 10,1,3) ,\bar\Sigma ( 10,1,3),{\rm except}\\
\;\;\;\;\;{\rm  \, for \, color \, singlets}
\end{array} $ & $ M_{PS}  $ 
& -421/5  & -51  & -65 \\
\hline
$\begin{array}{l}
\Phi(15,3,1),\\
\Phi(15,1,3)
\end{array} $ & $ M_R^2 M_X / M_{PS}^2  $ 
& -291/5  & -51  & -48 \\
\hline
$\begin{array}{l}
{\rm  \, color \; singlets \; from}\\
\Sigma(\bar 10,1,3) ,\bar\Sigma ( 10,1,3),{\rm except}\\
\;\;\;\;\;{\rm  for \; doubly -charged }, \delta^{++},\bar\delta^{--}
\end{array} $ & $ M_R  $ 
&  -153/5 & -21  & -24 \\
\hline
$\begin{array}{l}
{\rm  \,doubly-charged  }\\
\;\;\;\; \delta^{++},\bar\delta^{--}
\end{array} $ & $ 10 \, M_{PS}^2/M_X  $ 
& -33  & -21   & -24 \\
\hline
$\begin{array}{l}
\Phi(10,2,2),
\Phi(\bar 10,2,2)\\
{\rm  \, color \; octets \; from}\\
\Phi(15,1,1)
\end{array} $ & $ M_R^2/M_{PS}  $ 
& -141/5  & -21  &-24  \\
\hline
\end{tabular}
\end{center}
\caption{Particles that run and their approximate masses for case {\bf
B-1}\label{run}}
\end{table}

The scales in Table \ref{run} are in correct order provided
 $M_{PS}^3 > M_R^2 M_X$ and $M_R M_X > 10 M_{PS}^2$ .
It can immediately be seen that the enormous $b_i$ coefficients do not
allow for intermediate scales, there are simply too many states
running, and this is confirmed by a straightforward calculation.
Figure 1. shows the scales $M_X$, $M_{PS}$ and $M_R$ as functions 
of the unification constant, 
allowing for the errors in the value of $\alpha_3 = 0.117 \pm 0.002$.

\begin{figure}
\includegraphics{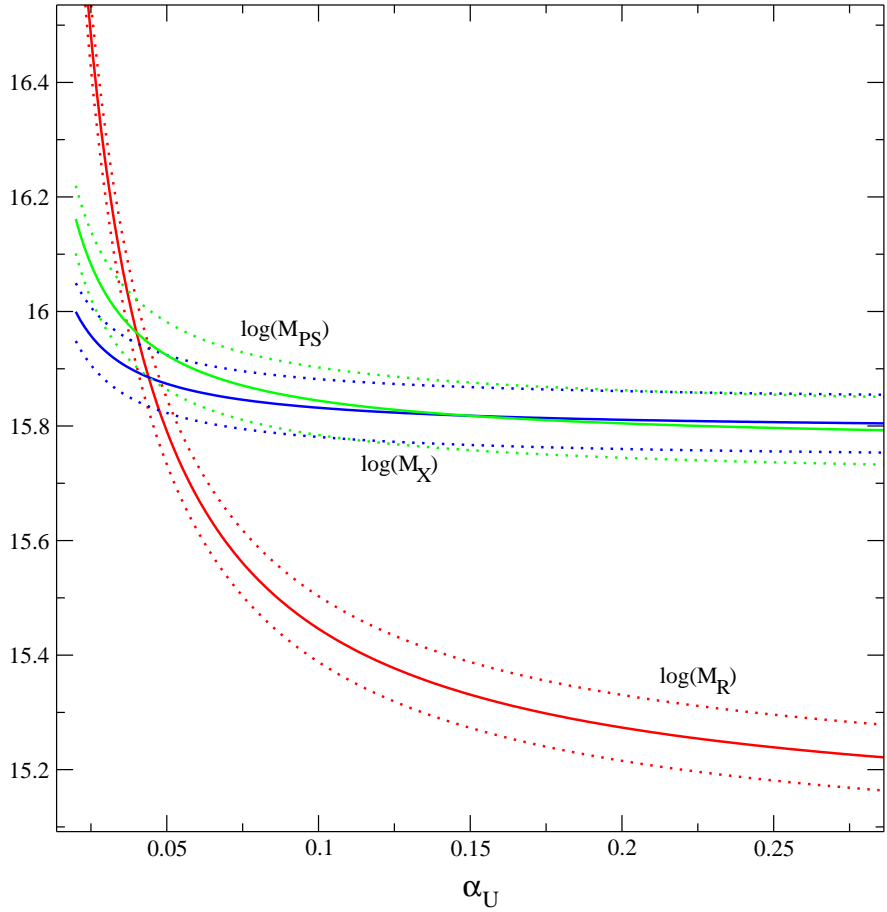}
\caption{$M_X$, $M_{PS}$ and $M_R$ as functions 
of the unification constant, for case {\bf B-1}. Dotted lines are
results for lowest and highest values of $\alpha_3$}.
\end{figure}

 If $M_{PS}^3 < M_R^2 M_X$, or if we consider case {\bf 2},
the same conclusion is reached: only the usual single-step breaking
is possible.

Strictly speaking, the complete spectrum that we present 
allows for a more profound analysis of the unification 
of the gauge couplings. One need not assume the order 
of magnitude equality of the masses in a fixed multiplet, 
but incorporate precisely the impact of every individual 
eigenstate mass. This may be worth doing but is beyond the 
scope of the present paper. 

\section{Summary and outlook}

In this work we addressed a computational (technical) investigation 
of the minimal renormalizable supersymmetric $SO(10)$ 
theory, such as a detailed study of the symmetry breaking 
and the calculation of the mass spectrum. These calculations 
are necessary for a detailed study of the phenomenological 
manifestations of this theory, namely  unification of the couplings, 
fermion masses, proton decay, leptogenesis, flavor violating processes, 
that will be performed in the sequel of this work (part 2).

In order to illustrate the calculations, we considered the question of
whether intermediate (gauge) scales are permitted in this theory,
reaching a negative conclusion. 
(Note however that it is possible that 
some particles of the theory can 
accidentally turn out to be light). 
A more detailed study, 
where the size of the threshold correction will be quantified, 
will be performed in part 2. 
Also, we studied the 
conditions imposed by the minimal fine tuning, 
that is needed to obtain the correct MSSM spectrum and that further 
reduces the number of free parameters. 
We demonstrated that the composition of the light 
higgs particles $H_u$ and $H_d$ is fixed in terms of the 
fundamental parameters of the theory. 

A possible use of our results is for example the issue of the 
nature of the see-saw mechanism, i.e. whether it is of type I 
or type II \cite{Lazarides:1980nt,Mohapatra:1980yp}. A simple 
and natural way to have a type I see-saw in the theory is to 
break the left-right symmetry at the GUT scale with the SU(2)$_R$ 
symmetry breaking scale much smaller \cite{Chang:fu}. This amounts 
to $p\gg\sigma ,\omega$, but it enters into the conflict with the 
unification constraints as discussed before, and even more important, 
creates more danger for the $d=5$ proton decay by lowering the masses 
of some colour triplets. 

On the other hand, a type II see-saw can be achieved by having a 
small mass of the lefthanded triplet through a judicious choice 
(fine-tuning) of the parameter $x$. Again, this can create conflict 
with the gauge coupling unification and ought to be checked. It is 
probably most natural to have both type I and type II compete on 
equal grounds. 

In summary, the minimal $SO(10)$ theory is a very constrained theory, 
with few free parameters. In this work, we prepared the necessary tools 
to explore this theory in detail.  We are not yet in position to assess 
which are its detailed predictions and as a matter of fact we cannot 
exclude that this theory will eventually fail. In view of several
experimental and theoretical considerations (e.g., on 
neutrino and fermion masses) we 
believe that such an exploration is largely worth the effort.

\section{Acknowledgments}
The work of A.M., G.S.\ and B.B.\ is supported by 
CDCHT-ULA (Project C-1073-01-05-A), 
EEC (TMR contracts ERBFMRX-CT960090 and
HPRN-CT-2000-00152) and the Ministry of Education, Science 
and Sport of the Republic of Slovenia, respectively. A.M.\ and F.V.\ thank 
ICTP and INFN exchange program (F.V.), for hospitality during the course 
of this work. We thank Charanjit S. Aulakh for innumerable discussions 
and cross-checking of many of our results. 

\section{Note added}

When these calculations were completed, 
a new paper appeared, also addressing the study of 
the minimal $SO(10)$ model~\cite{Fukuyama:2004xs}. 
There are some points of disagreement, 
and here we would like to argue in favor of 
our result. 
\begin{enumerate}
\item In their table 1, there are two 
$2\times 2$ matrices,  
whereas we have no such matrices. However, their existence 
would imply that $\Sigma$ mixes 
with itself (and similarly for $\bar\Sigma$),  while 
there is not such a coupling in the theory.\newline
\item Their eq.(4.2)
describes a $4\times 4$ mass matrix of the particles with SM numbers
(3,2,1/6), whereas we have a $3\times 3$ mass matrix, and a decoupled 
state, precisely, the component of $\Sigma$ with PS numbers $(15,2,2)$.
Our explanation of the existence of a 
decoupled state is the following one: 
the absence of a self coupling of $\Sigma$ (and $\bar\Sigma$) 
sets to zero one of their entries; 
the absence of the other couplings is due to 
the selection rule on $T_R$.\newline

\item After their eq.(5.7), it is argued that 
one should set $\alpha\neq \bar{\alpha}$,
if one wants to avoid the equality $Y_u=Y_d$.
Indeed, they note that when $\alpha = \bar{\alpha}$, 
the components of the vector that represent $H_u$  
are the same of the one of $H_d$.
However, the fields remain different, since 
the role of ${\Sigma}$ and $\bar{\Sigma}$ gets 
exchanged when going from $H_u$ to $H_d$.

\end{enumerate}
Similar computations have been performed by C.S. Aulakh and A.~Girdhar, 
to appear soon.

\appendix
\section{Decomposition of the SO(10) representations}
\label{one}

\subsection{Conventions}

Duality in $SO(2n)$ is defined as:
\beq
\Sigma^{\rm d}_{a_1..a_n} = \frac{- i^n}{n!} 
\epsilon_{a_1..a_n,b_1..b_n} \Sigma_{b_1..b_n}
\eeq

$SO(10)$ indices are labeled by latin subscripts, and
  is decomposed  so that 
 $i= 1..4$ is reserved for $SO(4)$  and $i=5..10$ for $SO(6)$. 

The color states
 in the fundamental {\bf 6} of $SO(6)$ are given by indices:
\beq
{\bf 3}_{-2/3} \begin{array}{cc} 
r : & 5 + i 6 \\
b: & 7 + i 8 \\
g: & 9 + i 0 
\end{array}  \eeq
with  $B-L = -2/3$. 
Then the 2-index {\bf 15} of $SU(4)$ is represented by:

\beq   
\begin{array}{lll}
{\bf 1}_0 & r\bar r + b\bar b +  g\bar g &=  [56 + 78 + 90] \\
\\
{\bf 3}_{4/3} & \bar r \bar g &= [59 - 60 - i50 - i69]\\
{\bf \bar 3}_{-4/3} & r g &= [59 - 60 + i50 + i69]\\
\\
{\bf 8}_0 & r\bar r - b \bar b = [ 56 - 78]
 \end{array}
\eeq
The 3-index {\bf 10}:
\beq   
\begin{array}{lll}
{\bf 1}_{-2} & r b g & = [579 -689 -670 -580 + i(679 + 589 + 570 - 680)]\\
\\
{\bf 3}_{-2/3} & (r\bar r + g\bar g) b & = -2 i[567 + 790 +i 568 + i 890]\\
\\
{\bf 6}_{2/3} & (-r \bar r + g\bar g)\bar b & = [568 - 890 - i 790 + i 567]
\end{array}
\eeq
The ${\bf \bar 10}$ is obtained by conjugation of above. The 4-index
${\bf 15}$ and the 5-index ${\bf 6}$ are obtained using dualization, from
the 2- and 1-index. 
$SU(2)_R$ doublets are given by:

\beq
\begin{array}{ll}
T_{3R}= +1/2 , \; T_{3L} = +1/2 : & [-1 + i 2] \\ 
T_{3R}= -1/2 , \; T_{3L} = +1/2  : & [3 + i 4] \\ 
\end{array}
\eeq

\beq
\begin{array}{ll}
T_{3R}= +1/2 , \; T_{3L} = -1/2 : & [ 3 - i 4] \\ 
T_{3R}= -1/2 , \; T_{3L} = -1/2  : & [1 + i 2] \\ 
\end{array}
\eeq

And 2-index ${\bf 3 + 1}$:
\beq
\begin{array}{ll}
T_{3R} = +1 :&  [14 + 23 + i (13 - 24)]\\ 
T_{3R} = 0 : &  [12 + 34]\\
T_{3R} = -1 :&  [14 + 23 - i (13 - 24)]\\
\\
T_{3R} = 0 : & [ 12 - 34]\\
\end{array}
\eeq
The 3-index doublets are dual to the 1-index. 

\subsection{States}

Standard Model states $\Phi_J,\, \Sigma_J,\, \bar{\Sigma_J}, H_J$ 
are combinations ( from here on capital indices label states) 

\beq
\Phi_J = c_J^{abcd} \Phi_{abcd} 
\eeq
and so on. One color representative $(7 + i 8)$
 is chosen for each color multiplet and
 the $T_{3L}=0$ for left-handed triplets.

Using indices for the states is not very practical, so for
 identification purposes SM states  
are labeled with a subindex:

$$SU(3),SU(2)_L , SU(2)_R ^{\;T_{3R}}  $$
 with a shorthand notation for
$T_{3R}$
So,  e.g. the fields that get a
vev would be  labeled:
$$   
\Omega = \Phi_{113^0} \quad \Sigma =\Sigma_{113^{-}} 
\quad \bar\Sigma=\bar\Sigma_{113^{+}}
$$

Ambiguities: The singlet in $\Phi(1,1,1)$ is called $\Phi_P$,
 while that in $\Phi(1,1,15)$ is called $\Phi_A$.  There are two 
$\Phi_{322^{\pm }}$, one in $(2,2,6)$, called $\Phi_{I^{\pm}}$ 
and one in $(2,2,10)$
called $\Phi_{II^{\pm }}$. 
Color singlets like $\Phi_{122}$ in $(2,2,10)$ are
distinguished from those in $(2,2,\bar{10})$ by calling them $\Phi_{\bar
122}$.
All states are normalized canonically in the kinetic term.

\subsection{Standard Model decomposition}

We give the  Standard Model states in terms of $SO(10)$ indices, labeled  by 
$SU(3),SU(2)_L , SU(2)_R ^{\;T_{3R}}  $ in Tables \ref{decomphi1}-\ref{decomsigma}. For $SU(3)$  states, 
only one color combination is given,
 and for $SU(2)_L$ triplets only the $T_{3L}=0$.  
 
States of $\Sigma$ ($\bar\Sigma$) are defined already self-dual 
(anti-self dual), but the dual part is not written in the table, 
for shortness, nor is the normalization factor included. States 
formed by the (antisymmetrized) linear combination 
of $N$ components of $\Phi_{ijkl}$
will be normalized by a factor $1/\sqrt{ 4! \, N}$. Similarly, 
states from $\Sigma_{ijklm}$ ($\bar \Sigma_{ijklm}$ ) 
will be normalized by $1/\sqrt{5 ! N}$,
where $N$ counts the field components and their dual (anti-dual) 
parts. States from $H_i$ have a $1/\sqrt{N}$ factor. 

The above states $\Phi$ are normalized canonically in a sense that 
the Kahler potential for them is $K=\Phi^\dagger\Phi$. It 
is easy to show that this corresponds to the original SO(10) 
Kahler

\begin{equation}
K={1\over 4!}\Phi_{ijkl}^\dagger\Phi_{ijkl}+
{1\over 5!}\Sigma_{ijklm}^\dagger\Sigma_{ijklm}+
{1\over 5!}\overline{\Sigma}_{ijklm}^\dagger\overline{\Sigma}_{ijklm}+
H_i^\dagger H_i
\end{equation}

\begin{table}
\begin{tabular}{|l|l|l|}
\hline
$G_{422}$ & State  in $\Phi$ & SO(10) indices \\
\hline

1,1,1  &

111= $P$  &$ [1,2,3,4]$ \\ 

\hline

15, 1,1 &
111= $A$  & $ [5,6,7,8] + [5,6,9,0] + [7,8,9,0]$\\ 

& 311 &  $ -[5,7,8,9] + [6,7,8,0] + i [6,7,8,9] + i [5,7,8,0] $\\ 

& $\bar 3$11  & $ -[5,7,8,9] +[6,7,8,0] - i[6,7,8,9] - i[5,7,8,0]
$\\ 

& 811 & $[7,8,9,0] - [5,6,9,0]$\\ 

\hline 

 15,1,3

& 113$^{-}$  &$ ( -i [1,3,5,6] -i [1,3,7,8] -i [1,3,9,0])$\\
            & &     $\;\;\; - (1,3 \to 2,4)
                    + i (1,3 \to 1,4) + i (1,3 \to 2,3)  $\\ 

& 113$^0$   &$([1,2,5,6] + [1,2,7,8] + [1,2,9,0]) + (1,2 \to
3,4))$\\

& 113$^{+}$   &$ ( i[1,3,5,6] + i[1,3,7,8] + i[1,3,9,0])$\\
            & &     $\;\;\; - (1,3 \to 2,4)
                    - i (1,3 \to 1,4) - i (1,3 \to 2,3)  $\\ 

&  313$^{-}$   &$ ( -i [1,3,5,9] +i [1,3,6,0] - [1,3,5,0] -  [1,3,6,9])$\\
            & &     $\;\;\;  - (1,3 \to 2,4)
                    + i (1,3 \to 1,4) + i (1,3 \to 2,3)  $\\

&  313$^0$   &$ ( [1,2,5,9] - [1,2,6,0] -i [1,2,5,0] - i [1,2,6,9])$\\
            & &     $\;\;\;
                           + (1,2 \to 3,4) $\\ 

&  313$^{+}$  &$( i[1,3,5,9] - i[1,3,6,0] + [1,3,5,0] + [1,3,6,9])$\\
            & &     $\;\;\;
                    - (1,3 \to 2,4)
                    - i (1,3 \to 1,4) - i (1,3 \to 2,3) $\\

& $\bar 3$13$^{-}$   &$( -i [1,3,5,9] +i [1,3,6,0]  + 
[1,3,5,0] + [1,3,6,9])$\\
            & &     $\;\;\;
                    - (1,3 \to 2,4)
                    + i (1,3 \to 1,4) + i (1,3 \to 2,3) $\\

& $\bar 3$13$^0$     &$( [1,2,5,9] - [1,2,6,0] + i [1,2,5,0] + i [1,2,6,9])$\\
            & &     $\;\;\;
                           + (1,2 \to 3,4) $\\

&  $\bar 3$13$^{+}$   &$ ( i [1,3,5,9] - i [1,3,6,0] - [1,3,5,0] - 
[1,3,6,9])$\\
            & &     $\;\;\;
                    - (1,3 \to 2,4)
                    - i (1,3 \to 1,4) - i (1,3 \to 2,3) $\\

& 813$^{-}$      &$ ( -i [1,3,5,6] + i [1,3,7,8]) - (1,3 \to 2,4)$\\
            & &     $\;\;\;\;
                    + i(1,3 \to 1,4) + i (1,3 \to 2,3) $\\

& 813$^0$  &$ ([1,2,5,6] - [1,2,7,8]) + (1,2 \to 3,4)$\\

& 813$^{+}$     &$ ( i [1,3,5,6] - i[1,3,7,8]) - (1,3 \to 2,4)$\\
            & &     $\;\;\;\;\;\;\;\;\;\;\;\;
                    - i (1,3 \to 1,4) - i (1,3 \to 2,3) $\\ 
      \hline           

 15,3,1

& 131   & $( [1,2,5,6] + [1,2,7,8] + [1,2,9,0]) - (1,2 \to
3,4)$\\ 

& 331    &$( [1,2,5,9] - [1,2,6,0] + i [1,2,5,0] + i [1,2,6,9)]$\\
            & &     $\;\;\;
                           - (1,2 \to 3,4 )$\\ 
               
& 831    &$([1,2,5,6] - [1,2,7,8]) - (1,2 \to 3,4)$\\ 

\hline

6,2,2

& 322$^+ = I^+$   &$ (- [2,3,4,7] - i [1,3,4,7]) + i (7\to  8 )$\\ 

& 322$^- =I^-$   &$ (+[1,2,4,7] - i [1,2,3,7])+ i (7\to  8 )$\\ 
    
& $\bar 3$22$^+ =\bar I^+$   &$ (- [2,3,4,7] - i [1,3,4,7])- i (7\to  8)$\\ 
         
& $\bar 3$22$^- = \bar I^-$  &$ (+[1,2,4,7] - i [1,2,3,7])- i (7\to  8
)$\\ 

 \hline   
\end{tabular}
\caption{Decomposition of states in the {\bf 210} representation.\label{decomphi1}}
\end{table}

\begin{table}
\begin{tabular}{|l|l|l|}
\hline
 $G_{422}$ & State  in $\Phi$ & SO(10) indices \\

  \hline 
10,2,2

& 122$^{+}$  &$ (-[1,5,7,9]+ [1,6,8,9] + [1,5,8,0] + [1,6,7,0]$\\
            & &     $\;\;\;
             -i [1,5,7,0] -i [1,5,8,9] -i [1,6,7,9] +i [1,6,8,0]) -i (1
\to 2) $\\

&  122$^{-}$   &$ (+[3,5,7,9]- [3,6,8,9] - [3,5,8,0] - [3,6,7,0]$\\
            & &     $\;\;\;
            + i [3,5,7,0]+ i[3,5,8,9] + i[3,6,7,9] - i[3,6,8,0]) +i (3
\to 4)$\\

&  322$^+ = {II}^+$   &$ (-[1,5,6,8] - [1,8,9,0] +i[1,5,6,7]
+i [1,7,9,0]) - i (1 \to 2 )$\\ 
             
&  322$^-= {II}^-$ &$ (+[3,5,6,8] +[3,8,9,0] -i[3,5,6,7]
-i[3,7,9,0])+i (3 \to 4)$\\

&  622$^{+}$ &$ (-[1,5,6,8] + [1,8,9,0]- i [1,5,6,7] + i [1,7,9,0])
-i (1 \to 2)$\\

&  622$^{-}$ &$ (+[3,5,6,8] - [3,8,9,0]+ i [3,5,6,7]- i [3,7,9,0] )
+i (3 \to 4 )$\\

  \hline

$\overline{10}$,2,2

& 122$^{+}$  &$ (-[1,5,7,9]+ [1,6,8,9] + [1,5,8,0] + [1,6,7,0]$\\
            & &     $\;\;\;
            + i[1,5,7,0]+ i[1,5,8,9] +i [1,6,7,9] - i[1,6,8,0]) -i (1
\to 2) $\\

&  122$^{-}$   &$ (+[3,5,7,9]- [3,6,8,9] - [3,5,8,0] - [3,6,7,0]$\\
            & &     $\;\;\;
            - i[3,5,7,0] -i [3,5,8,9] -i [3,6,7,9] +i [3,6,8,0]) +
i (3 \to 4)$\\

&  $\bar 3$22$^+= \overline{II}^+$   &$ (-[1,5,6,8] - [1,8,9,0] -i[1,5,6,7]
-i[1,7,9,0])- i (1 \to 2)$\\ 
             
&  $\bar 3$22$^- = \overline{II}^-$ &$ (+[3,5,6,8] +[3,8,9,0] +i[3,5,6,7]
+i[3,7,9,0])+i (3 \to 4)$\\

&  $\bar{6}$22$^{+}$ &$ (+[1,5,6,8] - [1,8,9,0]- i [1,5,6,7] + i
[1,7,9,0]) -i (1 \to 2)$\\

&  $\bar{6}$22$^{-}$ &$ (-[3,5,6,8] + [3,8,9,0]+ i [3,5,6,7] - i
[3,7,9,0]) +i (3 \to 4)$\\

\hline

\end{tabular}
\caption{Decomposition of states in the {\bf 210} representation. \label{decomphi2}}
\end{table}


\begin{table}
\begin{tabular}{|l|l|r|}
\hline
 $G_{422}$ & State  in $H$ & SO(10) indices \\ \hline

1,2,2 & 122$^{+}$     &   $ -[1] + i[2]$ \\ 
& 122$^{-}$     &   $   [3] + i[4]$ \\ 
\hline
6,1,1 & 311    &   $   [7] + i[8]$ \\ 
&$\bar{3}$11     &   $  [7] - i[8]$ \\  
\hline

\end{tabular}
\caption{Decomposition of states in the {\bf 10} representation.\label{decomh}}
\end{table}

\begin{table}
\begin{tabular}{|l|l|l|}
\hline 
 $G_{422}$ & State  in $\Sigma$ & SO(10) indices \\

\hline 

 6,1,1

& 311     & $ [1,2,3,4,7] + i[1,2,3,4,8] $ \\
& $\bar 3$11     &   $ [1,2,3,4,7] - i[1,2,3,4,8]$ \\

\hline

10,3,1

& 131           & $  ([1,2,5,7,9] -[1,2,6,8,9] 
              - [1,2,5,8,0] - [1,2,6,7,0]   $\\
            & &     $\;\;\;
           + i[1,2,5,7,0] + i[1,2,5,8,9]  
               +i[1,2,6,7,9] -i[1,2,6,8,0])   $\\
            & &     $\;\;\;\;\;\;\;\;\;\
                           - (1,2 \to 3,4) 
                                             $ \\ 

 & 331         &     $ ([1,2,5,6,8] + [1,2,8,9,0]
              -i[1,2,5,6,7] -i[1,2,7,9,0]) $\\
            & &     $\;\;\;  - (1,2 \to 3,4)
                                             $ \\

 & 631   &   $  
             ([1,2,5,6,8] -[1,2,8,9,0]+i[1,2,5,6,7] -i[1,2,7,9,0])  $\\
            & &     $\;\;\;\;\;\;\;\;\;\;  - (1,2 \to 3,4) 
                                             $ \\

\hline 

$\overline{10}$,1,3
 &  $\bar 1$13$^{-}$      & $ -i([1,3,5,7,9] -[1,3,6,8,9] 
              - [1,3,5,8,0] - [1,3,6,7,0]   $\\
            & &     $\;\;\;
           - i[1,3,5,7,0] - i[1,3,5,8,9]  
               -i[1,3,6,7,9] +i[1,3,6,8,0])  $\\
            & &     $\;\;\;- (1,3 \to 2,4)
                    + i \{1,3 \to 1,4\} + i \{1,3 \to 2,3\} $\\ 
 & $\bar 1$13$^0$     & $  ([1,2,5,7,9] -[1,2,6,8,9] 
              - [1,2,5,8,0] - [1,2,6,7,0]   $\\
            & &     $\;\;\;
           - i[1,2,5,7,0] - i[1,2,5,8,9]  
               -i[1,2,6,7,9] +i[1,2,6,8,0])   $\\
            & &     $\;\;\;
                           + (1,2 \to 3,4) 
                                             $ \\

 &  $\bar 1$13$^{+}$  & $ ( i[1,3,5,7,9] -i[1,3,6,8,9] 
              - i[1,3,5,8,0] - i[1,3,6,7,0]   $\\
            & &     $\;\;\;
           +[1,3,5,7,0] + [1,3,5,8,9] + 
                [1,3,6,7,9] - [1,3,6,8,0])  $\\
            & &     $\;\;\;- (1,3 \to 2,4)
                    - i (1,3 \to 1,4) - i (1,3 \to 2,3) $\\ 

 &   $\bar{3}$13$^{-}$     &   $ (-i [1,3,5,6,8] -i [1,3,8,9,0]
              + [1,3,5,6,7] + [1,3,7,9,0]) $\\
            & &     $\;\;\;- (1,3 \to 2,4)
                    + i (1,3 \to 1,4) + i (1,3 \to 2,3)$\\

 &  $\bar{3}$13$^0$      &     $ ([1,2,5,6,8] + [1,2,8,9,0]
              +i[1,2,5,6,7] +i[1,2,7,9,0]) $\\
            & &     $\;\;\;  + (1,2 \to 3,4)
                                             $ \\

 &   $\bar{3}$13$^{+}$    &   $ (i[1,3,5,6,8] + i[1,3,8,9,0]
              -[1,3,5,6,7] -[1,3,7,9,0]) $\\
            & &     $\;\;\;- (1,3 \to 2,4)
                    - i (1,3 \to 1,4) - i (1,3 \to 2,3) $\\ 
            
 &  $\bar 6$13$^{-}$    &      $  
             (-i [1,3,5,6,8] +i[1,3,8,9,0]- [1,3,5,6,7] +[1,3,7,9,0])  $\\
            & &     $\;\;\;- (1,3 \to 2,4)
                    + i (1,3 \to 1,4) + i (1,3 \to 2,3) $\\

 &  $\bar 6$13$^0$    &   $  
             ([1,2,5,6,8] -[1,2,8,9,0]-i[1,2,5,6,7] +i[1,2,7,9,0])  $\\
            & &     $\;\;\;  + (1,2 \to 3,4) 
                                             $ \\

 &  $\bar 6$13$^{+}$    &   $  
            (i[1,3,5,6,8] -i[1,3,8,9,0]+[1,3,5,6,7] -[1,3,7,9,0])  $\\
            & &     $\;\;\;- (1,3 \to 2,4)
                    - i (1,3 \to 1,4) - i (1,3 \to 2,3) $\\ 

\hline
 
 15,2,2

 &  122$^{+}$     &   $ 
               (-[1,5,6,7,8] - [1,5,6,9,0] - [1,7,8,9,0]) $\\
            & &     $\;\;\;  -i (1 \to 2)$\\ 

 &  122$^{-}$     &   $ 
               (+[3,5,6,7,8] + [3,5,6,9,0] + [3,7,8,9,0]) $\\
            & &     $\;\;\;  +i\{3 \to 4\}$\\ 

 &  322$^{+}$     & $-[1,6,7,8,9] -  [1,5,7,8,0]
                      -i[1,5,7,8,9] + i[1,6,7,8,0] $\\
            & &     $\;\;\;\;\;\;\;\;\;\;  -i (1 \to 2)$\\

 &  322$^{-}$    & $(+[3,6,7,8,9] + [3,5,7,8,0]
                      +i[3,5,7,8,9] - i[3,6,7,8,0]) $\\
            & &     $\;\;\;  +i(3 \to 4)$\\ 

 &  $\bar 3$22$^{+}$     & $(+[1,6,7,8,9] +  [1,5,7,8,0]
                -i[1,5,7,8,9] + i[1,6,7,8,0] )$\\
            & &     $\;\;\;  -i (1 \to 2)$\\

 &   $\bar 3$22$^{-}$    & $(-[3,6,7,8,9] -  [3,5,7,8,0]
                +i[3,5,7,8,9] - i[3,6,7,8,0]) $\\
            & &     $\;\;\;  - i (3 \to 4)$\\  

 &  822$^{+}$    &  $
                 (-[1,7,8,9,0] + [1,5,6,9,0])  -i(1 \to 2)$\\  
 
 &   822$^{-}$    & $
                (+[3,7,8,9,0] - [3,5,6,9,0])+i (3 \to 4)$\\  
 \hline   
\end{tabular}
\caption{Decomposition of states in the {\bf 126} representation\label{decomsigma}}
\end{table}

\section{Calculation of the spectrum}
\label{two}

\subsection{Symmetry breaking}

 The Higgs superpotential is:
\begin{eqnarray}
W_H &=& \frac{m_\Phi}{4!}\Phi_{ijkl}\Phi_{ijkl} +
\frac{\lambda}{4!}\Phi_{ijkl}\Phi_{klmn}\Phi_{mnij} \nonumber \\
&+ &\frac{m_\Sigma}{5!}\Sigma_{ijklm}\overline\Sigma_{ijklm} +
\frac{\eta}{4!}\Phi_{ijkl}\Sigma_{ijmno}\overline\Sigma_{klmno}\nonumber \\
&+&m_H H_i H_i + \frac{1}{4!} \Phi_{ijkl} H_m (\alpha
\Sigma_{ijklm} + \bar\alpha \overline{\Sigma}_{ijklm})
\label{Asuperpot}
\end{eqnarray}

Using the above conventions, the Standard Model singlet fields that
get a VEV are

\begin{eqnarray}
\langle \Phi_{1234}\rangle = p \; ;\quad 
\langle \Phi_{5678}\rangle  =\langle \Phi_{5690}\rangle 
=\Phi_{7890}\rangle  = a\nonumber \\
\langle \Phi_{1256}\rangle = \langle \Phi_{1278}\rangle 
=\langle \Phi_{1290}\rangle =
  \langle \Phi_{3456}\rangle  = \langle
\Phi_{3478}\rangle  = \langle \Phi_{3490}\rangle  = \omega
\nonumber \\
\langle\Sigma_{a+1,b+3,c+5,d+7,e+9}\rangle
=\frac{1}{2^{5/2}} (i)^{(-a-b+c+d+e)} \, \sigma \nonumber
\\
\langle\bar \Sigma_{a+1,b+3,c+5,d+7,e+9}\rangle
=\frac{1}{2^{5/2}} (-i)^{(-a-b+c+d+e)} \, \sigma
\end{eqnarray}
\noindent with $a,b,c,d,e,$ running for 0 to 1.
 The superpotential for this fields is then calculated to be
\begin{eqnarray}
W_H&=&m_\Phi\left(p^2+3a^2+6\omega^2\right)
+2 \lambda\left(a^3+3p\omega^2+6a\omega^2\right)\nonumber\\
&+&m_\Sigma \sigma\overline\sigma+\eta\sigma\overline\sigma\left(
p+3a -6 \omega\right)\; .
\label{vevspot}
\end{eqnarray}

Patterns of symmetry breaking are given in the text. 
We give a brief summary in Table \ref{sb} , specifying 
the Pseudo-Goldstone Bosons (PGB) when they exist.
The states that get their masses only after supersymmetry 
breaking can be found in our mass matrices  below.

F-term contribution to the masses are calculated by pieces, defining:

\begin{eqnarray}
\label{H}
({\cal H})_{a_1,a_2,a_3,a_4}&=&\frac{1}{6}\left[\langle\Phi
\rangle_{a_1,a_2,i,j}\Phi_{i,j,a_3,a_4}+\{a.s\}\right]\\ 
\label{Hs}
({\cal H}_{\sigma})_{a_1,a_2,a_3,a_4}&=&\frac{1}{6}\left[\langle\Sigma
\rangle_{a_1,a_2,i,j,k}\bar\Sigma_{i,j,k,a_3,a_4}+\{ a.s\}\right]\\ 
\label{J}
({\cal J})_{a_1,a_2,a_3,a_4,a_5}&=&\frac{1}{2}\left[
\frac{1}{10}\left(\langle\Phi\rangle_{a_1,a_2,i,j}
\Sigma_{i,j,a_3,a_4,a_5}+\{a.s\}\right)+\{dual\}\right]\\
\label{Js}
({\cal J}_\sigma)_{a_1,a_2,a_3,a_4,a_5}&=&\frac{1}{2}\left[\frac{1}{10}
\left(\langle\Sigma\rangle_{i,j,a_3,a_4,a_5}\Phi_{i,j,a_1,a_2}+\{a.s\}
\right)+\{dual\}\right]\\
\label{Ks}
({\cal K}_\sigma)_{a_1,a_2,a_3,a_4}&=&\langle\Sigma
\rangle_{a_1,a_2,a_3,a_4,i}H_i\\
\label{K}
({\cal K})_{a_1,a_2,a_3,a_4,a_5}&=&\frac{1}{2}\left[ \frac{1}{5}
\left(\langle\Phi\rangle_{a_1,a_2,a_3,a_4}H_{a_5}+ \{ a.s\}\right)
+\{dual\}\right]
\end{eqnarray}

(and similar for $\bar\Sigma$ ). Then, omitting $SO(10)$ indices

\begin{eqnarray}
F_\Phi &=& 2 m_\Phi \Phi + 6\lambda {\cal H} + \eta 
({\cal H}_{\bar\sigma} + {\cal H}_\sigma) + \alpha 
{\cal K}_\sigma + \bar \alpha {\cal K}_{\bar\sigma} \label{Fp}
\nonumber\\
F_{\bar\Sigma} &=& m_\Sigma \Sigma + 5\eta {\cal J} + 
5\eta {\cal J}_\sigma + 5\bar\alpha {\cal K}\label{Fs} 
\end{eqnarray}

To get the F-terms for the states, 
the combinations 
\beq
F_{\Phi_J} = \bar c_J^{abcd} F_{\Phi_{abcd}}
\label{fcomb}
\eeq
are found, where $\bar c_J$ are the coefficients of $\bar\Phi_J =\bar
c_J^{abcd}\Phi_{abcd}$. Note that the mass terms are of 
the form $\bar\Phi\Phi$ and of course in the case of SM singlets 
$\Phi=\bar\Phi$. Results are given in Tables \ref{F-phi} and \ref{F-sigma}

\begin{table}
\begin{tabular}{|l|r|r|r|r|}
\hline
  $F_\Phi$ & $6({\cal H}_p + {\cal H}_a)$  & $ 6 {\cal H}_\omega $ 
&$ {\cal H}_\sigma , {\cal H}_{\bar{\sigma}} $ &
${\cal K}_\sigma, {\cal K}_{\bar{\sigma}}$\\
\hline & & & & \\ 

 $F_{\Phi_P}$ &    &  $ 2\sqrt{6} \omega \, \Phi_\Omega$   &
              $  i(\bar\sigma \, \Sigma_{113^{-}} - \sigma \, 
\bar\Sigma_{113^+})
 $ & \\
                                                      
\hline

 $F_{\Phi_A}$ & $ 4 a \, \Phi_A$  & $ 4\sqrt{2} \omega \, \Phi_\Omega $  & 
       $ \sqrt{3}i(\bar\sigma \, \Sigma_{113^-}- \sigma \, 
\bar\Sigma_{113^+})$ 
                     &      \\ & & & & \\

 $F_{\Phi_{\bar 311}}$ & $2 a\,   \Phi_{311}$ & $ -2\sqrt{2} 
\omega\, \Phi_{313^0} $
 &                $ -\sqrt{2}\sigma \,\bar\Sigma_{313^+}$ &      
\\ & & & & \\

 $F_{\Phi_{ 311}}$ & $ 2 a\,  \Phi_{\bar 311}$ & $ -2\sqrt{2} 
\omega\, \Phi_{\bar
313^0}  $ &      $ -\sqrt{2}\bar \sigma \,\Sigma_{\bar 313^-}  $ &     
 \\ & & & & \\

$ F_{\Phi_{811}}$ & $ -2 a\, \Phi_{811}$ & $-2\sqrt{2}
\omega \, \Phi_{813^0} $  
                               &   &
\\
\hline

$F_{\Phi_{113^-} }$ & $2(p \, + 2 a \,)  \Phi_{113^+} $&  
              &$ i\sqrt{6}\bar \sigma \, \Sigma_{\bar 113^0}$   &    
 \\ 

  $F_{\Phi_\Omega}$& $ 2(p \, + 2 a \,) \Phi_{\Omega} $&$ 2\omega 
\,(\sqrt{6}\Phi_P + 2 
                  \sqrt{2}\Phi_A) $ &$ -\sqrt{6}i(\bar\sigma \,
\Sigma_{113^-} -\sigma \,
                        \bar\Sigma_{113^+})$  &      \\ 
               
  $F_{\Phi_{113^+} }$ & $ 2(p \, + 2 a \,)   \Phi_{113^-} $&  
              &$-i \sqrt{6} \sigma \, \bar \Sigma_{113^0} $   &     
\\ 

  $F_{\Phi_{\bar 313^-}}$ & $2(p \, + a \,) \Phi_{313^+} $ &$ 8 
\omega \, \Phi_{313^+} $   &   &     \\ 

  $F_{\Phi_{\bar313^0}}$& $ 2( p \, + a \,) \Phi_{313^0}   
$& $- 2\sqrt{2}\omega \,\, \Phi_{311} $  & $-2\sigma \,\bar 
\Sigma_{313^+}$   &     \\ 

  $F_{\Phi_{\bar 313^+}}$& $  2(p \, + a \,) \Phi_{313^-}  $
                 & $ -8\omega\,\Phi_{313^-} $ & 
              $ i\sqrt{2}\sigma \,(\bar\Sigma_{311} - i
\sqrt{2}\bar\Sigma_{313^0})$   
  & $i \sigma H_{311} $   \\ 

  $F_{\Phi_{ 313^-}}$ & $   2(p \, + a \,)  \Phi_{\bar 313^+} $ &
         $-8\omega\,\Phi_{\bar 313^+} $   & $ -i\sqrt{2}\bar\sigma 
\,(\Sigma_{\bar 311} + i\sqrt{2}\Sigma_{\bar 313^0})$  & $-i\bar
\sigma H_{\bar 311}  $  \\ 

 $F_{\Phi_{313^0}}$& $ 2(p \, + a \,) \Phi_{\bar 313^0}   
$& $-2\sqrt{2}\omega\Phi_{\bar 311} $  & $-2 \bar\sigma 
\Sigma_{\bar 313^-} $   &     \\ 

  $F_{\Phi_{ 313^+}}$& $2(p \, + a \,)  \Phi_{\bar 313^-}  $
                 & $ 8\omega\Phi_{\bar 313^-} $ & 
              $  $     &   \\ 

 $F_{\Phi_{813^-}}$& $ 2(p \, - a \,) \Phi_{813^+}  
$&  &   &     \\ 

 $F_{\Phi_{813^0}}$& $ 2( p \, - a \,) \Phi_{813^0} 
$&$-2\sqrt{2}\omega \,\Phi_{811}$
                                    &   &      \\ 

 $F_{\Phi_{813^+}}$& $ 2( p \, - a \,)  \Phi_{813^-}  $&  &   &     
\\ 

\hline 

 $F_{\Phi_{131} } $ &  $2(-p \, + 2 a \,)  \Phi_{131} $ &   &   &     
\\ 

$F_{\Phi_{\bar 331} } $  &$2(-p \, +  a \,)  \Phi_{331} $ &  &   &      
\\ 

 $F_{\Phi_{831} } $&  $2(-p \, - a\, )  \Phi_{831} $&  &   &   \\

\hline

 $F_{\Phi_{\bar I^-}   } $&    &$-2\omega \, (\Phi_{I^+} +
\sqrt{2}\Phi_{II^+})$
                            &$ -\sqrt{2}\bar\sigma \,
\Sigma_{322^-} $ &      \\ 

 $F_{\Phi_{\bar I^+}   } $&   & $2\omega \, (\Phi_{I^-} + 
\sqrt{2}\Phi_{II^-})$ 
 &   &      \\ 

 $F_{\Phi_{ I^-}   } $&    &$ 2\omega \, (\Phi_{\bar I^+} + 
\sqrt{2}\Phi_{\bar{II}^+}) $
                            &$  $ &      \\ 

 $F_{\Phi_{ I^+}   } $&    & $ -2\omega \, (\Phi_{\bar I^-} + 
\sqrt{2}\Phi_{\bar{II}^-})  $
                            & $-\sqrt{2}\sigma\bar
\Sigma_{\bar 322^+}  $ &      
\\
\hline

 $F_{\Phi_{\bar 122^-}} $&$ 6 a \, \Phi_{122^+}  $& $ -6 \omega \, 
\Phi_{122^+}  $          & $ \sqrt{6}\bar\sigma \,\Sigma_{122^-}$  
& $\bar\sigma \, H_{122^-}$
                                    \\ 

$F_{\Phi_{\bar 122^+}} $&$ 6 a \, \Phi_{122^-}  $& $ 6 \omega \, 
\Phi_{122^-}  $ 
         &  &
                                    \\ 

 $F_{\Phi_{\overline{II}^{-{1/2}}}} $&$ 2 a \, \Phi_{II^+}  
 $& $-2\omega \, (\Phi_{II^+} +
 \sqrt{2}\Phi_{I^+})$ & $-2 \bar\sigma \, \Sigma_{322^-}$   
&      \\ 

 $F_{\Phi_{\overline{II}^+}} $&$ 2 a \, \Phi_{II^-}   
$& $ 2\omega \, (\Phi_{II^-}                      
+ \sqrt{2}\Phi_{I^-})$ &   &      \\ 

 $F_{\Phi_{\bar 622^-}} $&$ -2 a \, \Phi_{622^+}  $&$ 2 
\omega \, \Phi_{622^+}  $ 
 &   &      \\ 

 $F_{\Phi_{\bar 622^+}} $&$ -2 a \, \Phi_{622^-}  $&$ -2 
\omega \, \Phi_{622^-}  $ 
 &   &      \\ 

\hline 

 $F_{\Phi_{ 122^-}} $&$ 6 a \, \Phi_{\bar 122^+}  $& $ 6
\omega \Phi_{\bar 122^+}  $          & $ $  & $ $
                                    \\ 

 $F_{\Phi_{ 122^+}} $&$ 6 a \, \Phi_{\bar 122^-}  $
& $-6\omega \Phi_{\bar 122^-}    $         & 
$-\sqrt{6}\sigma \bar\Sigma_{122^+} $ & $-\sigma H_{122^+} $
                                    \\ 

 $F_{\Phi_{ II^-}} $&$ 2 a \, \Phi_{\overline{II}^+}   $
& $2\omega \, (\Phi_{\overline{II}^+} +
 \sqrt{2}\Phi_{\bar I^+})  $ & $ $   &      \\ 

 $F_{\Phi_{ II^+}} $&$ 2 a \, \Phi_{\overline{II}^-}   $& $ -
2\omega \, (\Phi_{\overline{II}^-} +
 \sqrt{2}\Phi_{\bar I^-}) $ & $-2\sigma\bar\Sigma_{\bar 322^+}$  
&      \\ 

$F_{\Phi_{ 622^-}} $&$ -2 a \, \Phi_{\bar 622^+}  $&$ -2 \omega \, 
\Phi_{\bar 622^+}    $ 
 &   &      \\ 

$F_{\Phi_{ 622^+}} $&$ -2 a \, \Phi_{\bar 622^-}  $&$ 2 \omega \, 
\Phi_{\bar 622^-}  $ 
 &   &      \\ 

\hline
\end{tabular}
\caption{F-term contribution for states in $\Phi$.\label{F-phi}}
\end{table}

\begin{table}
\begin{tabular}{|l|r|r|r|r|}
\hline
$F_{\bar\Sigma}$ &$ 5 ({\cal J}_p + {\cal J}_a)$ & $5 {\cal J}_\omega$ &
 $5 {\cal J}_\sigma$ & $5 ({\cal K}_p + {\cal K}_a + {\cal K}_\omega)$\\

\hline

$F_{\bar\Sigma_{\bar 311}}$ &   &  &  & $\sqrt{\frac{1}{2}} 
(p \, -a \,)\, H_{311}$  
 \\ 

 $F_{\bar\Sigma_{311}}$ &   & $ -2i\sqrt{2} \omega \, 
\Sigma_{\bar 313^0}$ & $ \sqrt{2}i\sigma \, \Phi_{\bar 313^+} $ 
& $\sqrt{\frac{1}{2}}( p \, + a \,)\, H_{\bar 311}$ 
 \\ 

\hline

$F_{\bar\Sigma_{\bar 131}}$ & $ (-p \, + 3 a \,) \Sigma_{131}     $ & & & 
 \\ 

$F_{\bar\Sigma_{\bar 331}}$ & $ (-p \, + a \,)\Sigma_{331}     $& & & 
 \\ 

$F_{\bar\Sigma_{\bar 631}}$ & $ (-p \, - a \,)\Sigma_{631}    $& & &  
\\
\hline

$F_{\bar\Sigma_{113^-}}$ & $ (p \, + 3 a \,) \Sigma_{\bar 113^+}  $& 
$ 6 \omega \, \Sigma_{\bar 113^+}     $&  &  
 \\ 

$F_{\bar\Sigma_{113^0}}$ & $(p \, + 3 a \,) \Sigma_{\bar 113^0} $&  & 
           $ -i\sqrt{6}\sigma \, \Phi_{113^+} $ &  
\\ 

$F_{\bar\Sigma_{113^+}}$ & $ (p \, + 3a \,) \Sigma_{\bar 113^-}$&
                $ -6 \omega \, \Sigma_{\bar 113^-}$ & $-i\sigma \,
(\Phi_P + \sqrt{3}\Phi_A - \sqrt{6} \Phi_\Omega)   $ &      
\\ 

$F_{\bar\Sigma_{313^-}}$ & $ (p \, + a \,)\Sigma_{\bar 313^+}$&
                 $ 2 \omega \, \Sigma_{\bar 313^+}$ &  &  
\\ 

$F_{\bar\Sigma_{313^0}}$ & $ (p \, + a \,)\Sigma_{\bar 313^0} $& 
                $ 2\sqrt{2}i\omega \, \Sigma_{\bar 311} $ &
              $ 2\sigma \, \Phi_{\bar 313^+}$ & $ 2i\omega \, H_{\bar 311} $ 
 \\ 

$F_{\bar\Sigma_{313^+}}$ & $(p \, + a \,) \Sigma_{\bar 313^-} $& 
                 $ -2 \omega \,  \Sigma_{\bar 313^-} $ &
                $-\sqrt{2}\sigma \,(\Phi_{\bar 311} + \sqrt{2}\Phi_
{\bar 313^0})$
                &   
\\ 

$F_{\bar\Sigma_{613^-}}$ & $ (p \, - a \,) \Sigma_{\bar 613^+}     $&
                 $ -2\omega \,  \Sigma_{\bar 613^+}$&    &    
 \\ 

$F_{\bar\Sigma_{613^0}}$ & $(p \, - a \,)  \Sigma_{\bar 613^0}     $&  & &  
\\ 

$F_{\bar\Sigma_{613^+}}$ & $ (p \, - a \,) \Sigma_{\bar 613^-}     $
                   & $ 2 \omega \, \Sigma_{\bar 613^-}     $ &  &     \\ 

\hline

$F_{\bar\Sigma_{122^-}}$ & $2 a \, \Sigma_{122^+}$&
          $2 \omega \, \Sigma_{122^+}$&  & $\sqrt{\frac{3}{2}} 
(a\, +\omega \,) H_{122^+} $
 \\ 

$F_{\bar\Sigma_{122^+}}$ & $ 2 a \,\Sigma_{122^-}$&  
   $ -2 \omega \, \Sigma_{122^-}$&
    $\sqrt{6}\sigma \,\Phi_{122^+}$  & $\sqrt{\frac{3}{2}} (a \, -
\omega \,) H_{122^-} $  
 \\ 

$F_{\bar\Sigma_{\bar 322^-}}$ & $ a \,\Sigma_{322^+}$& 
       $ 3 \omega \, \Sigma_{322^+}$&  & \\ 

$F_{\bar\Sigma_{\bar 322^+}}$ & $a \, \Sigma_{322^-} $&
   $ -3 \omega \, \Sigma_{322^-} $& 
  $ -\sqrt{2}\sigma(\Phi_{I^+} + \sqrt{2}\Phi_{II^+})  $ & \\ 

$F_{\bar\Sigma_{322^-}}$ & $a \, \Sigma_{\bar 322^+} $& 
      $ -\omega \,  \Sigma_{\bar 322^+} $&  &   \\ 

$F_{\bar\Sigma_{322^+}}$ & $ a \,\Sigma_{\bar 322^-} $&
     $ \omega \,  \Sigma_{\bar 322^-} $& $  $ &   \\ 

$F_{\bar\Sigma_{822^-}}$ & $ -a \,\Sigma_{822^+} $& 
   $-\omega \,  \Sigma_{822^+} $&  &  \\ 

$F_{\bar\Sigma_{822^+}}$ & $-a \, \Sigma_{822^-} $& 
   $  \omega \, \Sigma_{822^-} $&   &  \\

\hline   
\end{tabular}
\caption{F-term contribution for states in $\Sigma$.\label{F-sigma}}
\end{table}


\subsection{Masses}

We give masses for eigenstates, or the mixing matrices, for the
most general pattern of symmetry breaking. States/matrices are
identified by the hypercharge $Y/2$.

 \subsubsection{Unmixed states}
\begin{center}
\begin{tabular}{|l|l|l|}
\hline
Y/2 & State & Mass \\
\hline
 (-5/3, 5/3) & $\Phi_{\bar 313^-}, \Phi_{313^+}   $ & $ 2 
(m_\Phi + \lambda (p + a + 4 \omega)) $\\
{ (-1 ,1 )}  & $ \Phi_{813^-},\Phi_{813^+}     $ & $  2 
( m_\Phi + \lambda (p - a))$ \\
{ (0 )}  & $ \Phi_{131}   $ & $ 2 (m_\Phi - \lambda(p - 2 a))$\\
{ ( -2/3, 2/3)}  & $ \Phi_{331}, \Phi_{\bar331}   $ & $ 2 (m_\Phi - 
\lambda(p - a)) $\\
{ (0 )} & $\Phi_{831}   $ & $ 2 (m_\Phi - \lambda(p + a))$\\
{ ( -3/2,3/2)}  & $ \Phi_{\bar122^+}, \Phi_{122^-} $ & $ 2 ( m_\Phi + 
3\lambda (a + \omega))$\\
{ ( -1/6,1/6)} & $ \Phi_{ 622^-}, \Phi_{\bar 622^+} $ & $ 2 (m_\Phi -
\lambda ( a + \omega))$\\
{ ( -5/6,5/6)} & $ \Phi_{\bar 622^-}, \Phi_{622^+} $ & $ 2 (m_\Phi -
\lambda ( a - \omega)) $\\
\hline 
{ (-1,1 )} & $ \Sigma_{131}, \bar \Sigma_{\bar131}  $ & $  
m_\Sigma-\eta(p - 3 a)     $\\
{ (-1/3,1/3 )} & $ \Sigma_{331}, \bar \Sigma_{\bar331}  $ & $ 
m_\Sigma-\eta (p - a)      $\\
{ ( 1/3,-1/3)} & $ \Sigma_{631}, \bar \Sigma_{\bar 631}  $ & $ 
 m_\Sigma-\eta (p + a)     $\\
{ ( 2,-2)} & $ \Sigma_{\bar 113^+}, \bar \Sigma_{113^-}  $ & $ 
m_\Sigma+\eta (p + 3a +6\omega)     $\\
{ ( 4/3,-4/3)} & $ \Sigma_{\bar 313^+}, \bar \Sigma_{313^-}  $ & $ 
m_\Sigma+\eta (p + a + 2\omega)     $\\
{ ( 2/3,-2/3)} & $ \Sigma_{\bar 613^+}, \bar \Sigma_{613^-}  $ & $ 
m_\Sigma+\eta (p - a - 2\omega)     $\\
{ ( -1/3,1/3)} & $ \Sigma_{\bar 613^0}, \bar \Sigma_{613^0}  $ & $ 
m_\Sigma+\eta (p - a )     $\\
{ ( -4/3,4/3)} & $ \Sigma_{\bar 613^-}, \bar \Sigma_{613^+}  $ & $ 
m_\Sigma+\eta (p - a + 2\omega)     $\\
{ ( 7/6,-7/6)} & $ \Sigma_{322^+}, \bar \Sigma_{\bar322^-}  $ & $  
m_\Sigma +\eta (a + 3\omega) $\\  
{ ( -1/6,1/6)} & $ \Sigma_{\bar 322^+}, \bar \Sigma_{322^-}  $ & $  
m_\Sigma +\eta (a -\omega)   $\\
{ ( -7/6,7/6)} & $ \Sigma_{\bar 322^-}, \bar \Sigma_{322^+}  $ & $  
m_\Sigma +\eta (a +\omega)   $\\
{ ( 1/2,-1/2)} & $ \Sigma_{822^+}, \bar \Sigma_{822^-}  $ & $  
m_\Sigma -\eta (a +\omega)   $\\
{ ( -1/2,1/2)} & $ \Sigma_{822^-}, \bar \Sigma_{822^+}  $ & $
m_\Sigma -\eta (a -\omega)   $\\
\hline
\end{tabular}
\end{center}

\subsubsection{Mixing matrices}

 For mixed states, these are the fermion mass matrices:

$$ \frac{\partial^2 W}{\partial \overline\varphi_i \partial \varphi_j}  $$

\noindent{\bf Y/2=(0,0)}

\noindent$\varphi=(\Phi_P,\Phi_A, \Phi_\Omega, \Sigma_{113^-}, 
\overline\Sigma_{113^+})   $

\noindent$\overline\varphi=(\Phi_P,\Phi_A, \Phi_\Omega,
\overline\Sigma_{113^+}, \Sigma_{113^-})   $

\beq
\left(
\begin{array}{ccccc}
2m_\Phi & 0                  & \lambda 2\sqrt{6} \omega & 
i\eta \bar\sigma      & -i\eta \sigma       \\
0  & 2(m_\Phi + 2 \lambda a) & \lambda 4 \sqrt{2}\omega & 
i\eta \sqrt{3}\bar\sigma & -i\eta\sqrt{3}\sigma \\
\lambda 2\sqrt{6} \omega & \lambda 4 \sqrt{2}\omega & 2(m_\Phi + 
\lambda(p+2a))&-i\eta\sqrt{6}\bar\sigma &i \eta\sqrt{6}\sigma \\
-i\eta \sigma & -i\eta\sqrt{3}\sigma & i\eta\sqrt{6} \sigma
 & m_\Sigma+\eta(p + 3a - 6\omega)& 0 \\
i\eta \bar \sigma & i \eta\sqrt{3}\bar\sigma & -i \eta \sqrt{6}
\bar\sigma & 0
 &  m_\Sigma+\eta(p + 3a - 6\omega)
\end{array}
\right)
\eeq

\noindent {\bf Y/2=(-2/3,2/3)}

\noindent $\varphi= (\Phi_{\bar 311} ,\Phi_{\bar 313^0}, \Sigma_{\bar 313^-})$

\noindent$\overline\varphi=( \Phi_{ 311} ,\Phi_{ 313^0}, 
\overline\Sigma_{ 313^+})$

\beq
\left(
\begin{array}{ccc}
2 (m_\Phi + \lambda a) & -\lambda 2\sqrt{2}\omega & -\eta\sqrt{2}\bar\sigma \\
-\lambda 2\sqrt{2}\omega & 2(m_\Phi + \lambda(p+a)) & -\eta 2\bar\sigma \\
-\eta \sqrt{2}\sigma & -\eta 2 \sigma & m_\Sigma+ \eta(p + a - 2 \omega) 
\end{array}
\right)
\eeq

\noindent {\bf Y/2 = (1,-1)}

\noindent $\varphi=( \Phi_{113^+}, \Sigma_{\bar 113^0}) $

\noindent$\overline\varphi=( \Phi_{113^-},  \overline\Sigma_{ 113^0}) $

\beq
\left(
\begin{array}{cc}
2 (m_\Phi + \lambda (p + 2 a)) & i\eta \sqrt{6}\bar\sigma \\
-i \eta \sqrt{6}\sigma & m_\Sigma+ \eta (p + 3 a)
\end{array}
\right)
\eeq

\noindent {\bf Y/2 =(0,0)   }

\noindent $\varphi = ( \Phi_{811} , \Phi_{813^0}  )   $

\noindent$\overline\varphi = ( \Phi_{811} , \Phi_{813^0}  )   $

\beq
\left(
\begin{array}{cc}
2(m_\Phi - \lambda a) & -\lambda 2\sqrt{2} \omega \\
-\lambda 2\sqrt{2} \omega & 2(m_\Phi +\lambda(p-a))
\end{array}
\right)
\eeq

\noindent {\bf Y/2=(1/6,-1/6)   }

\noindent $\varphi= (\Phi_{I^+}, \Phi_{II^+},\Sigma_{322^-}  )   $

\noindent$\overline\varphi= (\Phi_{\overline{I}^-}, 
\Phi_{\overline{II}^-},\overline\Sigma_{\bar 322^+}  )   $

\beq
\left(
\begin{array}{ccc}
2 ( m_\Phi - \lambda \omega) & -\lambda 2\sqrt{2} \omega & -
\eta \sqrt{2} \bar\sigma \\
- \lambda 2\sqrt{2} \omega & 2(m_\Phi + \lambda (a - \omega)) & -
\eta 2\bar\sigma \\
 -\eta \sqrt{2} \sigma & -\eta 2\sigma & m_\Sigma+ \eta (a - 3 \omega)
\end{array}
\right)
\eeq

\noindent {\bf Y/2 = (-5/6,5/6)   }

\noindent $\varphi = (\Phi_{I^-},  \Phi_{II^-}  )   $

\noindent$\overline\varphi = (\Phi_{\overline{I}^+},  
\Phi_{\overline{II}^+}  )   $

\beq
\left(
\begin{array}{cc}
2 ( m_\Phi +\lambda \omega) & \lambda 2\sqrt{2}\omega \\
\lambda 2\sqrt{2}\omega & 2(m_\Phi + \lambda (a + \omega))
\end{array}
\right)
\eeq

\noindent {\bf Y/2 = (-1/2,1/2)   }

\noindent $\varphi = ( H_{122^-} ,\Sigma_{122^-} , 
\overline\Sigma_{122^-} ,\Phi_{122^+} ) $

\noindent$\overline\varphi = ( H_{122^+} ,\overline\Sigma_{122^+} , 
\Sigma_{122^+} ,\Phi_{\bar122^-} ) $

\beq
\left(
\begin{array}{cccc}
2 m_H & \alpha \sqrt{\frac{3}{2}}(a - \omega) & \bar \alpha 
\sqrt{\frac{3}{2}}(a + \omega)  &\alpha \sigma \\
\bar \alpha \sqrt{\frac{3}{2}}(a - \omega) & m_\Sigma+ 2\eta 
(a -\omega) & 0 & \eta\sqrt{6}\sigma \\
\alpha \sqrt{\frac{3}{2}}(a + \omega) & 0 & m_\Sigma+ 2\eta 
(a +\omega) & 0 \\ 
-\bar\alpha\bar\sigma & -\eta \sqrt{6}\bar\sigma & 0 & -2
(m_\Phi + 3 \lambda (a -\omega)) 
\end{array}
\right)
\eeq

\noindent {\bf Y/2 = (1/3,-1/3)  }

\noindent $\varphi = (H_{\bar 311} , \Sigma_{\bar 311},  
\overline\Sigma_{\bar 311} ,\Sigma_{\bar 313^0} , \Phi_{\bar 313^+}  )  $

\noindent $\overline\varphi = (H_{ 311} , \overline\Sigma_{ 311},  
\Sigma_{ 311} ,\overline\Sigma_{ 313^0} , \Phi_{ 313^-}  )  $

\beq
\left(
\begin{array}{ccccc}
2 m_H & \frac{\alpha}{\sqrt{2}}(p + a) & \frac{\bar\alpha}{\sqrt{2}}(p - a) &
-\alpha 2 i \omega & \alpha i \sigma \\
\frac{\bar \alpha}{\sqrt{2}}(p + a) & m_\Sigma& 0 & -\eta 2 i 
\sqrt{2}\omega & \eta i\sqrt{2}\sigma \\
\frac{\alpha}{\sqrt{2}}(p - a) & 0 & m_\Sigma& 0 & 0\\
\bar\alpha 2i \omega & \eta 2 i \sqrt{2} \omega & 0 & m_\Sigma+\eta(p + a) &
\eta 2 \sigma \\
-\bar\alpha i \bar\sigma & -\eta i \sqrt{2}\bar\sigma & 0 & 
\eta 2 \bar\sigma & 2(m_\Phi +\lambda(p + a - 4 \omega) )
\end{array}
\right)
\eeq

\section{Properties of the mixing matrices}
\label{three}


\subsection{Standard Model vacuum}

In the Standard Model vacuum, we have

\beq
p =-\frac{m_\Phi}{\lambda} \frac{x (1-5x^2)}{(1 - x)^2} \;;\;
a=-\frac{m_\Phi}{\lambda} \frac{(1 - 2x - x^2)}{(1-x)} \;;\; 
\omega = - \frac{m_\Phi}{\lambda}x \;;\; 
\sigma^2 =\frac{2m_\Phi^2}{\eta\lambda} 
\frac{x (1 -3 x)(1 + x^2)}{(1 - x)^2} \;;
\label{A7vevs}
\eeq

\noindent
with

\beq 
-8 x^3 + 15 x^2 - 14 x + 3 = (x-1)^2 
\frac{\lambda m_\Sigma}{\eta m_\Phi}
\label{A7g}
\eeq

Eigenstates are listed in Tables I and II. We give here the mixing 
matrices for the SM singlets, color triplets and $SU(2)_L$ doublets 
for the SM vacua parametrized by $x$. Matrices are in units of $m_\Phi$.

\noindent {\bf Y/2 = 0}

\noindent $\varphi=(\Phi_P,\Phi_A, \Phi_\Omega, \Sigma_{113^-}, 
\overline\Sigma_{113^+})   $

\noindent$\overline\varphi=(\Phi_P,\Phi_A, \Phi_\Omega,\overline
\Sigma_{113^+}, \Sigma_{113^-})   $

\beq
\left(
\begin{array}{ccccc}
2 & 0                  & - 2\sqrt{6} x & i
\sqrt{\frac{\eta}{\lambda}} s(x)     & -i
\sqrt{\frac{\eta}{\lambda}} s(x)      \\
0  & -2\frac{2x^2 + 3x -1}{(x-1)} & - 4 
\sqrt{2}x & i\sqrt{\frac{\eta}{\lambda}} 
\sqrt{3} s(x) & -i\sqrt{\frac{\eta}{\lambda}}\sqrt{3}s(x) \\
- 2\sqrt{6} x & - 4 \sqrt{2}x & 2  
\frac{(1+x^2)(3x -1)}{(x-1)^2}&-i\sqrt{\frac{\eta}{\lambda}}
\sqrt{6} s(x) &i\sqrt{\frac{\eta}{\lambda}} \sqrt{6} s(x) \\
-i\sqrt{\frac{\eta}{\lambda}} s(x) & -i\sqrt{\frac{\eta}{\lambda}}
\sqrt{3}s(x) & i\sqrt{\frac{\eta}{\lambda}}\sqrt{6}s(x)
 &  0 & 0 \\
i\sqrt{\frac{\eta}{\lambda}} s(x) & i\sqrt{\frac{\eta}{\lambda}} 
\sqrt{3}s(x) & -i\sqrt{\frac{\eta}{\lambda}} \sqrt{6} s(x) & 0
 &  0
\end{array}
\right)
\eeq
where 
$$s(x) =\pm
 \frac{\sqrt{ 2 x (1 -3 x)(1 + x^2)}}{(x - 1)}  $$

\noindent {\bf Y/2 = (-1/2,1/2) }

\noindent $\varphi = ( H_{122^-} ,\Sigma_{122^-} , 
\overline\Sigma_{122^-} ,\Phi_{122^+} ) $

\noindent$\overline\varphi = ( H_{122^+} ,\overline\Sigma_{122^+} , 
\Sigma_{122^+} ,\Phi_{\bar122^-} ) $

\beq
\left(
\begin{array}{cccc}
2 m_H/m_\Phi &  \frac{\alpha}{\lambda} \sqrt{\frac{3}{2}}
\frac{1-3x}{x-1} &  \frac{\bar \alpha}{\lambda} 
\sqrt{\frac{3}{2}}\frac{-2 x^2- x +1}{x-1}  &  
\frac{\alpha}{\sqrt{\eta\lambda}} s(x) \\
 \frac{\bar \alpha}{\lambda} \sqrt{\frac{3}{2}}\frac{1-3x}{x-1}  
& -\frac{\eta}{\lambda}\frac{8x^3-9x^2+6x-1}{(x-1)^2}   & 0 
& \sqrt{6}\sqrt{\frac{\eta}{\lambda}} s(x) \\
 \frac{\alpha}{\lambda} \sqrt{\frac{3}{2}}\frac{-2 x^2- x +1}{x-1}   
& 0 & -\frac{\eta}{\lambda}\frac{12 x^3 -17 x^2+10 x -1}{(x-1)^2}  & 0 \\ 
-\frac{\bar\alpha}{\sqrt{\eta\lambda}} s(x) &  -
\sqrt{6}\sqrt{\frac{\eta}{\lambda}}  s(x) & 0 & -4\frac{1 - 4x}{x-1}
\end{array}
\right)
\eeq

\noindent {\bf Y/2 = (1/3,-1/3)  }

\noindent $\varphi = (H_{\bar 311} , \Sigma_{\bar 311},  
\overline\Sigma_{\bar 311} ,\Sigma_{\bar 313^0} , \Phi_{\bar 313^+}  )  $

\noindent $\overline\varphi = (H_{ 311} , \overline\Sigma_{ 311},  
\Sigma_{ 311} ,\overline\Sigma_{ 313^0} , \Phi_{ 313^-}  )  $

\beq
\left(
\begin{array}{ccccc}
2 m_H/m_\Phi & \frac{\alpha}{\lambda}\frac{1}{\sqrt{2}}
\frac{4x^3 -x^2+2x-1}{(x-1)^2} & \frac{\bar\alpha}{\lambda}
\frac{1}{\sqrt{2}}\frac{(1+x)(3x-1)(2x-1)}{(x-1)^2} &
 2 i\frac{\alpha}{\lambda} x &  i \frac{\alpha}{\sqrt{\eta\lambda}} s(x)  \\
\frac{\bar \alpha}{\lambda}\frac{1}{\sqrt{2}}
\frac{4x^3 -x^2+2x-1}{(x-1)^2} & -\frac{\eta}{\lambda}
\frac{8x^3-15x^2+14x-3}{(x-1)^2} & 0 &  2 i\sqrt{2}
\frac{\eta}{\lambda}x   &  i\sqrt{2}\sqrt{\frac{\eta}{\lambda}} s(x) \\
\frac{\alpha}{\lambda}\frac{1}{\sqrt{2}}\frac{(1+x)(3x-1)(2x-1)}{(x-1)^2}  
& 0 & -\frac{\eta}{\lambda}\frac{8x^3-15x^2+14x-3}{(x-1)^2}  & 0 & 0\\
- 2 i\frac{\bar\alpha}{\lambda} x & - 2 i \sqrt{2}
\frac{\eta}{\lambda} x  
& 0 & -2 \frac{\eta}{\lambda}\frac{2x^2-5x+1}{(x-1)^2} &
 2 \sqrt{\frac{\eta}{\lambda}} s(x) \\
- i \frac{\bar\alpha}{\sqrt{\eta\lambda}} s(x)  & - i \sqrt{2}
\sqrt{\frac{\eta}{\lambda}} s(x) & 0 &  2 
\sqrt{\frac{\eta}{\lambda}} s(x) & 8 x \frac{2 x^2 -2 x + 1}{(x-1)^2}
\end{array}
\right)
\eeq

\subsection{Explicit expression of the determinants}
There is some interest in having an expression of the 
determinant of the matrices.
One reason is that one wants that some determinants are not small.
Another reason is the following.
Let us consider the 1 loop running of 
$n$ particles with masses $m_1,m_2$ ... $m_n$, 
that mix among them through the $n\times n$ 
matrix ${\cal M}$. Since  
they have same beta function coefficient $b$, their contribution is
$b\log(m_1/M)+b\log(m_2/M)+$ .... $b\log(m_n/M)=
b\log(\prod_{i=1}^{n} m_i/M^n)$. Of course, the masses 
$m_i$ have to be positive numbers here 
and the zero modes have to be excluded. Thus one does not 
need exactly the determinant of the matrix,
but the square  root of 
${\cal M}_{n} {\cal M}_{n}^\dagger$, with the zero modes removed,
that we call `reduced determinant'. 
(In the case of a matrix $1\times 1$, 
one has simply to take the absolute value.) 
For matrices $n\times n$ with a single zero mode, 
namely $m_1=0$, the reduced determinant can 
be calculated either explicitly or by a simpler formula:
\begin{equation}
\prod_{i=2}^{n} m_i = 
\left[\, {\rm Det}'({\cal M}{\cal M}^\dagger)\, \right]^{1/2}= 
{\rm Abs}\left[\,
{{\rm Det}'({\cal M}) }\, /\, {\langle e|f\rangle} \, \right] 
\end{equation}
where the symbol ${\rm Det'}$ means $
{\rm Det}'({\cal M})=
\lim_{\epsilon\to 0} {\rm Det}({\cal M}+\epsilon\, 1\!\!\! 1)/\epsilon$.
At the denominator of the last expression 
we have $\langle e|f\rangle=
\sum_{i=1}^n e_i^* f_i$, namely 
the scalar product of the unit vectors  
of the left and right zero modes defined by 
${\cal M} f=0$ and ${\cal M}^\dagger e=0$. 

In the following, we show the determinants ``Det'' of the mass matrices  
and the reduced determinants ``Det$'$'' identified by their $G_{321}$
quantum numbers. 

\begin{equation}
{\rm Det}(8,1,0)=8\,m_{\Phi}^2\,
\frac{\left( x +1\right) \,\left( 2\,x -1\right) \,
    \left( x^3 + 6\,x^2 - 7\,x +2 \right) }{{\left( x -1\right) }^3}
\end{equation}

\begin{equation}   
{\rm Det}(3,1,-1/3)=-32\, m_{\Phi}^5\, 
\frac{{\alpha}\, {\bar\alpha}\,{\eta}^2}{\lambda^4}\, 
\frac{x\,
    \left( x+1 \right) \,{\left(  2\,x-1 \right) }^2\,
    {\left( 3\,x -1 \right) }^2\,
    p_{16} }{{\left( x -1\right) }^9\, p_3 p_5}
\end{equation}

\begin{equation}   
{\rm Det}'(1,1,1)=2\, m_{\Phi}
\frac{ \left( 3\,x -1\right) \,\left( x^2+1 \right)
- 3 \frac{\eta}{\lambda} {\left( x -1\right) }^2\,x}
{{\left( x-1 \right) }^2}
   \end{equation}

\begin{equation}
{\rm Det}'(3,2,-5/6)=-2\, m_{\Phi}\ 
\frac{ 3\,x^2  -2 x+1}{x-1}
\end{equation}

\begin{equation}   
{\rm Det}'(3,1,2/3)=-8\, m_{\Phi}^2 x\, 
\frac{\left( 3\,x -1\right) \,x\,
       \left( x^2+1 \right)  -
      \frac{\eta}{\lambda} 
\,\left( 3\,x^4 + 4\,x^2 - 4\,x + 1 
\right) }{{\left( x-1 \right) }^3}
 \end{equation}

\begin{equation}  
{\rm Det}'(3,2,1/6)=-4\,m_{\Phi}^2\,
\frac{ 2\,x\, {\left( x -1 \right) }^2\,
       \left( x^2 +1\right)  + 
      \frac{\eta}{\lambda} \left( 3\,x -1\right) \,
         \left( 2\,x^4 + x^2 - 2\,x + 1 \right) }{
{\left(  x-1 \right) }^3}
\end{equation}

\begin{equation}   
{\rm Det}'(1,1,0)=128\, m_{\Phi}^4\ \frac{\eta}{\lambda}\, 
\frac{x\,\left( x +1 \right) \,\left( 2\,x -1\right) \,
    \left( 3\,x -1\right) \,\left( x^2 +1 \right) \,
    \left( x^3 - 3\,x^2 + 2\,x  -1 \right) }{
    {\left( x-1 \right) }^5}
\end{equation}

\begin{equation}   
{\rm Det}'(1,2,1/2)=2\,m_{\Phi}^3\,
\frac{ 
 \frac{{\alpha}\, {\bar\alpha}}{\lambda^2}\,
       \left(          6\,{\left( x-1 \right) }^2\, p_{14}
+ \frac{\eta}{\lambda}\,x\,
            p_{15}      \right)  \ +
2\,{\left( \frac{\eta}{\lambda} \, p_{3}\, p_5 \right)^2}}{
    {\left(  x -1 \right) }^5\, p_3\, p_5 }
\end{equation}
The relevant polynomials are:

\begin{center}
\begin{tabular}{c|l}
\hline
$p_3$    & $12 x^3-17 x^2+10 x-1$ \\ \hline
$p_3'$ & $4 x^3-9x^2+9x-2$ \\ \hline
$p_5$    &  $9 x^5+20 x^4-32 x^3+21 x^2-7x +1$ \\ \hline
$p_{10}$ & $90 x^{10}-858 x^9+2009 x^8-3073 x^7+4479 x^6-$\\
& $5018 x^5+ 3618 x^4-1545 x^3+377 x^2-50 x+3$\\ \hline
$p_{14}$ &  $810 x^{14} - 1386 x^{13} + 10403 x^{12} - 30182 x^{11} +$ \\
& $61077\,x^{10} - 103524\,x^9 + 138678\,x^8 - 134068\,x^7 + $ \\
& $ 90165\,x^6 - 41308\,x^5 + 12678\,x^4 - 2562\,x^3 +$ \\
& $ 332\,x^2 - 26\,x +1$ \\ \hline
$p_{15}$ & $ (p_3\, p_3')^2 ( 3 x^3 - x^2 +  3 x -1)$ \\ \hline
$p_{16}$ & $1983\,x^{16} - 12829\,x^{15} + 28050\,x^{14} - 8886\,x^{13} -$  
 \\ 
& $ 100962\,x^{12} + 308738\,x^{11} - 540127\,x^{10} +  677679\,x^9 - $
 \\ 
& $ 644005\,x^8 + 465661\,x^7 -  252382\,x^6 + 100018\,x^5 - $ \\
& $28070\,x^4 + 5374\,x^3 - 677\,x^2 + 53\,x  -2 $ \\ \hline
\end{tabular}
\end{center}

Finally, we give the expressions of the (unnormalized) left and 
right zero-modes, useful to calculate the reduced determinants:
\begin{equation}
\left\{
\begin{array}{l}
f(1,1,1)=(-i\sqrt{6}\omega,\sigma)\\
e^*(1,1,1)=(i\sqrt{6}\omega,\sigma)  
\end{array}
\right. \ \ \ 
\left\{
\begin{array}{l}
f(3,2,-5/6)=(-\sqrt{2} x,x-1)\\
e^*(3,2,-5/6)=f(3,2,-5/6)
\end{array}
\right. \ \ \ 
\left\{
\begin{array}{l}
f(3,1,-2/3)= (- \sqrt{2} a, 2 \omega ,\sigma)\\
e^*(3,1,-2/3)=f(3,1,-2/3)
\end{array}
\right. 
\end{equation}

\begin{equation}
\left\{
\begin{array}{l}
f(3,2,-1/6)=((\omega-p)/\sqrt{2},\omega-a,\sigma)\\
e^*(3,2,-1/6)= f(3,2,-1/6) 
\end{array}
\right.\ \ \ 
\left\{
\begin{array}{l}
f(1,1,0)=(0,0,0,1,1)  \\
e^*(1,1,0)=f(1,1,0)
\end{array}
\right.
\end{equation}
The expressions for the zero modes of the doublet matrix
(=the light higgses) are given 
and discussed in the next Section.

\subsection{Arranging 2 light doublets}

In order to arrange the right spectrum in low energy theory,
one has to impose that the determinant of the mass matrix 
of the doublets vanishes in first approximation.
Since no coupling or mass of the theory 
is zero, this condition (referred as ``minimal fine-tuning'') 
reads as follows:

\begin{equation}
m_H=m_\Phi \frac{\alpha{\bar{\alpha}}}{2 \eta\lambda} 
\frac{p_{10}}{(x-1) p_3 p_5}
\end{equation}

\noindent
where the three polynomials of $x$, namely $p_3,p_5$ and $p_{10}$ 
are defined above.
{}From this condition, one finds 
the expressions of the zero-modes:

\begin{eqnarray}
H_d&\propto& 
\frac{2 p_5}{x-1} H_{122^-}
-\sqrt{6}\; \frac{\bar{\alpha}}{\eta} (3x-1) (x^3+5x-1) 
\Sigma_{122^-}-\sqrt{6}\; \frac{{\alpha}}{\eta} (2x-1) (x+1) 
\frac{p_5}{p_3}\overline{\Sigma}_{122^-} 
+ \bar{\alpha} \frac{\sigma}{m_\Phi} p_3'\Phi_{122^+}\;,\\
H_u&\propto& 
\frac{2 p_5}{x-1} H_{122^+}
-\sqrt{6}\; \frac{\alpha}{\eta} (3x-1) (x^3+5x-1) 
\overline{\Sigma}_{122^+}-\sqrt{6}\; \frac{\bar{\alpha}}{\eta} (2x-1) (x+1) 
\frac{p_5}{p_3}\Sigma_{122^+} 
- {\alpha} \frac{\sigma}{m_\Phi} p_3'\Phi_{122^-}
\end{eqnarray}

\noindent
(the proper normalization of $H_d$ is obtained in an obvious manner).
The components that matter for the coupling of $H_d$ to light 
particles are the first and the third one, while those for the 
coupling of $H_u$ are the first two.

It should be noted that the mixing between the 
10- and 126-light component depends just on two parameters, 
namely $x$ and $\alpha/\eta$. 
Furthermore, the parameter $\alpha/\eta$ always 
multiplies the coupling $Y_\Sigma$.
Finally, we illustrate 
the dependence on $x$ by giving a pair of examples
(from here on, we have in mind the case $|\alpha/\eta|\sim 1$).
If we ask that $H_u$ has a reduced $\overline{126}$ 
component, we are either at $x\sim 1/3$ or close
at one of the roots of $x^3+5x-1$--the real one is close to $0.2$. 
Convesely, it is interesting to note that 
there is a root of $p_3$ in the vicinity of $x=0.12$, and there 
$\overline{126}$ component of $H_d$ happens to be large.

\begin{table}[b]
\begin{center}
\begin{tabular}{|l|l|c|l|}
\hline
  & Vevs & $SO(10) \to $ & P.G.B.\\ \hline
\hline
1 & $p=a=w=\sigma=0$ & -- & -- \\ \hline
2 & \begin{tabular}{l}
$p=a=w=-m/3\lambda;$\\ $\sigma=0$ \end{tabular} 
& $SU(5)\times U(1)$ & -- \\ \hline
3 &\begin{tabular}{l} $p=a=w= -M/10\eta$;\\$ \sigma^2 = 
M(10\eta m - 3 \lambda M)/50 \eta^3$ \end{tabular} & $SU(5)$ & -- \\ \hline
4 &\begin{tabular}{l} $p=w=\sigma=0$ ;\\$ a=-m/\lambda $ \end{tabular}
&$ SU(3)_c\times SU(2)_L\times SU(2)_R\times U(1)_{B-L}$ &
\begin{tabular}{l}$(3 + \bar 3, 1, 3)$,\\$(3 +\bar 3, 3,1 )$
\end{tabular} \\ \hline 
5 &\begin{tabular}{l} $p=a=-w =-m/3\lambda$; \\$\sigma=0 $ \end{tabular}&
 $^fSU(5)\times U(1)$& -- \\ \hline
6 &\begin{tabular}{l} $ p = -3m/\lambda; a=-2 m/\lambda $ ;\\ 
$ w=\pm i m/\lambda ; \sigma =0 $\end{tabular} &
$ SU(3)_c\times SU(2)_L\times U(1)_R \times U(1)_{B-L} $ &
$(8,3,1) $ \\ \hline
7 &  eqs. (\ref{7vevs})-(\ref{7g}) & $ SU(3)_c\times 
SU(2)_L\times U(1)_Y $ & depend on $x$ \\ \hline
\hline
\end{tabular}
\caption{Patterns of symmetry breaking\label{sb}}

\end{center}
\end{table}


\begin{thebibliography}{3}

\bibitem{Aulakh:2003kg}
C.~S.~Aulakh, B.~Bajc, A.~Melfo, G.~Senjanovi\' c and F.~Vissani,
arXiv:hep-ph/0306242.

\bibitem{mc}
C.S.~Aulakh, R.N.~Mohapatra,
Phys.\ Rev.\ D {\bf 28} (1983) 217

\bibitem{Clark:ai}
T.~E.~Clark, T.~K.~Kuo and N.~Nakagawa,
Phys.\ Lett.\ B {\bf 115} (1982) 26.

\bibitem{Aulakh:1997ba}
C.S.~Aulakh, K.~Benakli, G.~Senjanovi\'c,
Phys.\ Rev.\ Lett.\  {\bf 79} (1997) 2188
[arXiv:hep-ph/9703434].

\bibitem{Aulakh:1998nn}
C.~S.~Aulakh, A.~Melfo and G.~Senjanovi\'c,
Phys.\ Rev.\ D {\bf 57} (1998) 4174
[arXiv:hep-ph/9707256].

\bibitem{Aulakh:1999cd}
C.S.~Aulakh, A.~Melfo, A.~Ra\v sin, G.~Senjanovi\' c,
Phys.\ Lett.\ B {\bf 459} (1999) 557
[hep-ph/9902409].

\bibitem{yana}
T.~Yanagida, proceedings of the {\em Workshop on Unified Theories 
and Baryon Number in the Universe}, Tsukuba, 1979, eds. 
A. Sawada, A. Sugamoto, KEK Report No. 79-18, Tsukuba.

\bibitem{glashow}
S.~Glashow, in {\em Quarks and Leptons, Carg\`ese 1979}, eds. 
M. L\' evy. et al., (Plenum, 1980, N ew York).

\bibitem{gmrs}
M.~Gell-Mann, P.~Ramond, R.~Slansky, proceedings of the
{\em Supergravity Stony Brook Workshop}, New York, 1979, 
eds. P. Van Niewenhuizen, D. Freeman (North-Holland, Amsterdam).

\bibitem{Mohapatra:1979ia}
R.~Mohapatra, G.~Senjanovi\' c,
Phys.Rev.Lett. {\bf 44} (1980) 912 

\bibitem{Babu:1992ia}
K.~Babu, R.~Mohapatra,
Phys.Rev.Lett. {\bf 70} (1993) 2845
[hep-ph/9209215].

\bibitem{Oda:1998na}
K.y.~Oda, E.~Takasugi, M.~Tanaka, M.~Yoshimura,
Phys.\ Rev.\ D {\bf 59} (1999) 055001
[hep-ph/9808241].

\bibitem{Brahmachari:1997cq}
B.~Brahmachari, R.N.~Mohapatra,
Phys.\ Rev.\ D {\bf 58} (1998) 015001
[hep-ph/9710371].

\bibitem{Bajc:2002iw}
B.~Bajc, G.~Senjanovi\' c, F.~Vissani,
Phys.\ Rev.\ Lett.\  {\bf 90} (2003) 051802[hep-ph/0210207] 
and hep-ph/0110310.

\bibitem{Matsuda:2001bg}
K.~Matsuda, Y.~Koide, T.~Fukuyama, H.~Nishiura,
Phys.\ Rev.\ D {\bf 65} (2002) 033008
[E.-ibid.\ D {\bf 65} (2002) 079904]
[hep-ph/0108202].

\bibitem{Goh:2003sy}
H.~S.~Goh, R.~N.~Mohapatra and S.~P.~Ng,
Phys.\ Lett.\ B {\bf 570} (2003) 215
[arXiv:hep-ph/0303055].

\bibitem{Goh:2003hf}
H.~S.~Goh, R.~N.~Mohapatra and S.~P.~Ng,
Phys.\ Rev.\ D {\bf 68} (2003) 115008
[arXiv:hep-ph/0308197].

\bibitem{c}
C.S.~Aulakh,
hep-ph/0210337.

\bibitem{Lee:1993rr}
D.~G.~Lee,
Phys.\ Rev.\ D {\bf 49} (1994) 1417.

\bibitem{charanaarti}
C.~S.~Aulakh and A.~Girdhar,
arXiv:hep-ph/0204097.



\bibitem{He:jw}
X.~G.~He and S.~Meljanac,
Phys.\ Rev.\ D {\bf 41} (1990) 1620.

\bibitem{Georgi:1979df}
H.~Georgi and C.~Jarlskog,
Phys.\ Lett.\ B {\bf 86} (1979) 297.

\bibitem{Sakai:1981pk}
N.~Sakai, T.~Yanagida,
Nucl.\ Phys.\ B {\bf 197} (1982) 533

\bibitem{Weinberg:1981wj}
S.~Weinberg,
Phys.\ Rev.\ D {\bf 26} (1982) 287

\bibitem{Hisano:1992jj}
J.~Hisano, H.~Murayama and T.~Yanagida,
Nucl.\ Phys.\ B {\bf 402} (1993) 46
[arXiv:hep-ph/9207279].

\bibitem{Lucas:1996bc}
V.~Lucas and S.~Raby,
Phys.\ Rev.\ D {\bf 55} (1997) 6986
[arXiv:hep-ph/9610293].

\bibitem{Goto:1998qg}
T.~Goto and T.~Nihei,
Phys.\ Rev.\ D {\bf 59} (1999) 115009
[arXiv:hep-ph/9808255].

\bibitem{Bajc:2002bv}
B.~Bajc, P.~Fileviez Perez and G.~Senjanovi\' c,
Phys.\ Rev.\ D {\bf 66} (2002) 075005
[arXiv:hep-ph/0204311].

\bibitem{Bajc:2002pg}
B.~Bajc, P.~Fileviez Perez and G.~Senjanovi\' c,
arXiv:hep-ph/0210374.

\bibitem{Emmanuel-Costa:2003pu}
D.~Emmanuel-Costa and S.~Wiesenfeldt,
Nucl.\ Phys.\ B {\bf 661} (2003) 62
[arXiv:hep-ph/0302272].

\bibitem{Murayama:2001ur}
H.~Murayama and A.~Pierce,
Phys.\ Rev.\ D {\bf 65} (2002) 055009
[arXiv:hep-ph/0108104].

\bibitem{Goh:2003nv}
H.~S.~Goh, R.~N.~Mohapatra, S.~Nasri and S.~P.~Ng,
arXiv:hep-ph/0311330.

\bibitem{Dimopoulos:1981yj}
S.~Dimopoulos, S.~Raby, F.~Wilczek,
Phys.\ Rev.\ D {\bf 24} (1981) 1681

\bibitem{Ibanez:yh}
L.E.~Ibanez, G.G.~Ross,
Phys.\ Lett.\ B {\bf 105} (1981) 439

\bibitem{Einhorn:1981sx}
M.B.~Einhorn, D.R.~Jones,
Nucl.\ Phys.\ B {\bf 196} (1982) 475

\bibitem{Marciano:1981un}
W.~Marciano, G.~Senjanovi\' c,
Phys.Rev.D {\bf 25} (1982) 3092

\bibitem{Lazarides:1980nt}
G.~Lazarides, Q.~Shafi and C.~Wetterich,
Nucl.\ Phys.\ B {\bf 181} (1981) 287.

\bibitem{Mohapatra:1980yp}
R.~N.~Mohapatra and G.~Senjanovi\' c,
Phys.\ Rev.\ D {\bf 23} (1981) 165.

\bibitem{Chang:fu}
D.~Chang, R.~N.~Mohapatra and M.~K.~Parida,
Phys.\ Rev.\ Lett.\  {\bf 52} (1984) 1072.

\bibitem{Fukuyama:2004xs}
T.~Fukuyama, A.~Ilakovac, T.~Kikuchi, S.~Meljanac and N.~Okada,
supersymmetric SO(10) GUT,''
arXiv:hep-ph/0401213.

\bibitem{Chacko:1998jz}
Z.~Chacko and R.~N.~Mohapatra,
Phys.\ Rev.\ D {\bf 59} (1999) 011702
[arXiv:hep-ph/9808458].

\end{thebibliography}
\end{document}